\documentstyle[epsf,epsfig,rotate,onecolumn]{mn}
\newcommand{\xist}{\mbox{\boldmath $\xi^\ast$}}
\newcommand{\be}{\begin{equation}}
\newcommand{\ee}{\end{equation}}
\newcommand{\ben}{\begin{eqnarray}}
\newcommand{\een}{\end{eqnarray}}

\newcommand{\no}{\noindent}

\newcommand{\bnab}{\mbox{\boldmath $\nabla$}}	
\newcommand{\btimes}{\mbox{\boldmath $\times$}}
\newcommand{\beq}{\begin{equation}}
\newcommand{\eeq}{\end{equation}}
\newcommand{\bde}{\begin{displaymath}}
\newcommand{\ede}{\end{displaymath}}

\newcommand{\bmin}[1]{\begin{minipage}[b]{#1\linewidth}}
\newcommand{\emin}{\end{minipage}}
\newcommand{\bfig}{\begin{figure}}
\newcommand{\efig}{\end{figure}}

\newif\ifAMStwofonts
\AMStwofontstrue

\ifoldfss
  \ifCUPmtlplainloaded \else
    \NewTextAlphabet{textbfit} {cmbxti10} {}
    \NewTextAlphabet{textbfss} {cmssbx10} {}
    \NewMathAlphabet{mathbfit} {cmbxti10} {} 
    \NewMathAlphabet{mathbfss} {cmssbx10} {} 
  \fi
  \ifAMStwofonts
    \ifCUPmtlplainloaded \else
      \NewSymbolFont{upmath} {eurm10}
      \NewSymbolFont{AMSa} {msam10}
      \NewMathSymbol{\upi}     {0}{upmath}{19}
      \NewMathSymbol{\umu}     {0}{upmath}{16}
      \NewMathSymbol{\upartial}{0}{upmath}{40}
      \NewMathSymbol{\leqslant}{3}{AMSa}{36}
      \NewMathSymbol{\geqslant}{3}{AMSa}{3E}

    \fi
  \fi
\fi 

\ifnfssone
  \newmathalphabet{\mathit}
  \addtoversion{normal}{\mathit}{cmr}{m}{it}
  \addtoversion{bold}{\mathit}{cmr}{bx}{it}

  \def\textbfss{\protect\txtbfss}
  
  \long\def\txtbfss#1{{\fontfamily{cmss}\fontseries{bx}\fontshape{n}%
    \selectfont #1}}
  \newmathalphabet{\mathbfit} 
  \addtoversion{normal}{\mathbfit}{cmr}{bx}{it}
  \addtoversion{bold}{\mathbfit}{cmr}{bx}{it}
  \newmathalphabet{\mathbfss} 
  \addtoversion{normal}{\mathbfss}{cmss}{bx}{n}
  \addtoversion{bold}{\mathbfss}{cmss}{bx}{n}
  \ifAMStwofonts
    \ifCUPmtlplainloaded \else
      \UseAMStwoboldmath
      \makeatletter
      \new@mathgroup\upmath@group
      \define@mathgroup\mv@normal\upmath@group{eur}{m}{n}
      \define@mathgroup\mv@bold\upmath@group{eur}{b}{n}
      \edef\UPM{\hexnumber\upmath@group}
      \new@mathgroup\amsa@group
      \define@mathgroup\mv@normal\amsa@group{msa}{m}{n}
      \define@mathgroup\mv@bold\amsa@group{msa}{m}{n}
      \edef\AMSa{\hexnumber\amsa@group}
      \makeatother
      \mathchardef\upi="0\UPM19
      \mathchardef\umu="0\UPM16
      \mathchardef\upartial="0\UPM40
      \mathchardef\leqslant="3\AMSa36
      \mathchardef\geqslant="3\AMSa3E
    \fi
  \fi
\fi 

\ifnfsstwo

  \def\textbfss{\protect\txtbfss}
  
  \long\def\txtbfss#1{{\fontfamily{cmss}\fontseries{bx}\fontshape{n}%
    \selectfont #1}}
  \DeclareMathAlphabet{\mathbfit}{OT1}{cmr}{bx}{it}
  \SetMathAlphabet\mathbfit{bold}{OT1}{cmr}{bx}{it}
  \DeclareMathAlphabet{\mathbfss}{OT1}{cmss}{bx}{n}
  \SetMathAlphabet\mathbfss{bold}{OT1}{cmss}{bx}{n}
  \ifAMStwofonts
    \ifCUPmtlplainloaded \else
      \DeclareSymbolFont{UPM}{U}{eur}{m}{n}
      \SetSymbolFont{UPM}{bold}{U}{eur}{b}{n}
      \DeclareSymbolFont{AMSa}{U}{msa}{m}{n}
      \DeclareMathSymbol{\upi}{0}{UPM}{"19}
      \DeclareMathSymbol{\umu}{0}{UPM}{"16}
      \DeclareMathSymbol{\upartial}{0}{UPM}{"40}
      \DeclareMathSymbol{\leqslant}{3}{AMSa}{"36}
      \DeclareMathSymbol{\geqslant}{3}{AMSa}{"3E}
    \fi
  \fi
\fi 

\ifCUPmtlplainloaded \else
  \ifAMStwofonts \else 
    \def\upi{\pi}
    \def\umu{\mu}
    \def\upartial{\partial}
  \fi
\fi

\title{ Bending instabilities in magnetized accretion discs}
\author[V. Agapitou, J. C. B. Papaloizou and C. Terquem]
       {Vasso Agapitou$^1$\thanks{V.Agapitou@qmw.ac.uk},
	 John C. B. Papaloizou $^1$ and 
       Caroline Terquem$^{2,1,3}$ \\
       $^1$ Astronomy Unit, School of Mathematical Sciences,
       Queen Mary \& Westfield College, Mile End Road, London E1 4NS, UK \\
       $^2$ Lick Observatory, University of California, Santa Cruz, 
       CA 95064, USA \\
       $^3$ Laboratoire d'Astrophysique, Universit\'e Joseph Fourier/CNRS, 
       BP 53, 38041 Grenoble Cedex 9, France}

\date{Received; Accepted}
\volume{000}
\pagerange{\pageref{firstpage}--\pageref{lastpage}}
\pubyear{1997}

\begin{document}

\maketitle

\label{firstpage}

\begin{abstract}

\no We study the global bending modes of a thin annular disc subject to both an
internally generated magnetic field and a magnetic field 
due to a dipole embedded in the central star 
with axis aligned with the disc rotation axis.
When there is a significant inner region of
the disc corotating with the star, we find spectra of unstable bending
modes. These may lead to elevation of the disc above the original symmetry
plane facilitating accretion along the magnetospheric field lines.
The resulting non-axisymmetric disc configuration may
 result in the creation of
hot spots on the stellar surface and the periodic photometric variations 
observed in many classical T Tauri stars (CTTS). Time--dependent behaviour
 may occur including the shadowing of the central source in magnetic accretors
 even when the  dipole and rotation axes are aligned.

\end{abstract}

\section{Introduction}

\noindent Situations in which a thin accretion disc is threaded by a
strong poloidal magnetic field are of interest in different
astrophysical contexts relating to accreting objects such as neutron
stars or young stellar objects.
This situation may result when the disc interacts with a
magnetic field produced by a dipole embedded in the central star
 such that the dipole field lines
penetrate the disc (Ghosh~\& Lamb~1978, Ghosh~\& Lamb~1979, Campbell~1987, Camenzind~1990,
K\"onigl~1991).
 
\noindent In addition open poloidal field lines may be
advected inwards by the accreting matter (Lubow,
Papaloizou~\& Pringle~1994, Reyez-Ruiz~\& Stepinski~1996,
 Agapitou~\& Papaloizou 1996) and be associated with a
centrifugally driven wind (see  K\"onigl~1993 for a review).
The observed correlation between mass accretion rate and mass outflow
rate in T Tauri stars (TTS)  supports the
idea of the accretion disc as the underlying source of the outflows emanating
from young stars (Cabrit et al.~1990).

\noindent  The short-term photometric variability 
evident in many CTTS has been
attributed to the presence of both dark and hot spots that cover a part
of the stellar surface (Bouvier~1994, Bouvier et al.~1995).
The latter have been  explained as arising from shocks formed close
to the stellar surface resulting from non-axisymmetric accretion along 
stellar magnetic field lines. Such magnetospheric accretion
is also invoked to explain the large infall velocities inferred from emission
 lines (Calvet~\& Hartmann~1992, Edwards et al.~1994, Hartmann, 
Hewett~\& Calvet~1994), the infrared colours of TTS 
(Kenyon, Yi~\& Hartmann~1996)
 and the outbursts of EX Lupi (Lehmann, Reipurth~\& Brandner~1995).  
Most of the variability 
of TTS cannot be accounted for unless there is some time--dependence in the 
magnetic field and(or) the accretion flow (Bouvier et al.~1995).
The presence of an accretion disc around many TTS seems also to be
linked to their low rotational velocities and the kind of activity
observed on the stellar surface (Edwards et al.~1993, Montmerle et al.~1993).

\noindent Under some conditions, the interaction of a strong stellar dipole field
 with a  surrounding low mass disc prevents accretion. This is because the
magnetic stresses exerted on the disc, external to the radius where the star
corotates with the local disc material which has near--Keplerian
rotation, act so as to transfer angular momentum to the disc and spin down
 the star. The accretion disc is then truncated at a radius where 
the viscous and magnetic torques  balance. 
Models of this type have been developed to account for the low
rotational velocities of TTS provided that the accretion disc is not
dissipated too early in the star's lifetime (Cameron~\& Campbell 1993, Yi~1994,
Ghosh~1995, Armitage~\& Clarke~1996). We note though that non--magnetized
stars could also experience spin-down while they accrete
mass. Paczynski~(1991)
 and Popham~\& Narayan~(1991) have shown that when a star rotates close to
its breakup speed the accreted specific angular momentum decreases
and it even attains negative values for large enough values of the stellar
rotation speed.

\noindent The situation of no or low mass accretion together with stellar
spin down occurs when the field is strong and the disc mass is low
giving rise to a small  viscous  angular momentum flux. 
In this paper we shall consider the converse situation when the disc is massive
enough to  penetrate through to the corotation region so that accretion becomes
possible. As in previous studies, we assume the existence of a  dipole field
embedded in the central star. The present observational evidence cannot
 rule out such a coherent field structure
 (Montmerle et al.~1994). There is evidence from numerical calculation
of non-linear stellar dynamos that a steady dipole mode is the most easily
excited one (Brandenburg, Tuominen~\& Moss~1989). The possibility of a fossil
field has also been proposed (Tayler~1987).

\noindent In this paper  we adopt the model of Spruit~\& Taam~(1990)
which is such that when the disc
is able to reach  interior to the corotation radius,
the inner parts  which contain field lines connected to the
central star,  corotate with it. In that
case, simple thin disc models in which the magnetic field plays an
important part in supporting the inner corotating region against
gravity can be constructed when the axis of the dipole and the disc
angular momentum axis are aligned (Spruit~\& Taam~1990, 1993). 
Considering the inner corotating
disc to be part of the magnetosphere, these models have the fastness
parameter, or ratio of stellar to inner differentially rotating disc
angular velocity, close to unity.  Magnetic support may also be
important in the outer, differentially rotating part of the disc if an
inwardly advected field becomes strong, as would be the case  in the 
presence of a strong wind removing most of the disc angular 
momentum (see for example K\"onigl~1989).

\noindent The importance of the issue of the stability of these disc
models has been stressed by Spruit~\& Taam~(1990) who considered the
role of interchange instabilities in enabling matter to migrate
inwards in the inner corotating region of the disc until direct
particle motion confined to equilibrium vacuum field lines becomes possible.
However, note that such particle motion along vacuum field lines
may not be representative of the plasma flow that may occur along field
lines because the plasma may in principle contain significant
currents that disturb the original vacuum field lines. A study of the possible
flow along field lines accordingly requires a full MHD treatment. The local
stability of the outer differentially rotating disc to interchange
modes has been considered by Spruit, Stehle~\& Papaloizou~(1995).

\noindent In this paper, we study the global stability of a thin
magnetized accretion disc to both axisymmetric and non-axisymmetric
disturbances perpendicular to its plane (bending modes). These are the
thin disc limit of modes with density perturbation having odd symmetry
with respect to reflection in the mid--plane, in contrast to the
interchange modes which have even symmetry. At equilibrium the disc is
permeated by both an internally produced poloidal magnetic field and
an external dipole field. We here limit consideration to the case when
the vertical component of the field in the inner disc does not change
sign. Bending modes are of potential interest because an instability
may lead to elevation of the disc mid-plane above the original
symmetry plane leading to facilitated motion along field lines connected to the
star as well as time--dependent shadowing of the central source. In the
context of neutron stars this was also pointed out by Spruit~\& Taam~(1990).
The  accretion
along magnetic field lines derived from a non-axisymmetric disc would result in
the production of hot spots on the stellar surface and  modulation of the
power output. This could  lead to a time--dependent accretion flow even in the
aligned dipole case and it may
account  for the irregular variability of CTTS without the need to invoke  a
variable magnetic field. Finally, non-axisymmetric modes with azimuthal mode
number $m=1$ are related to disc precession (Papaloizou~\& Terquem~1995)
and are of potential interest with regard to precessing jets.

\noindent In section~\ref{sec:equil} we describe the thin equilibrium
disc models. In section~\ref{sec:pert} we give the perturbation
equations for linear modes under the {\em ab initio} assumption of a 
razor--thin disc. We derive the local dispersion relation which indicates
instability in the inner corotating parts of the disc if they are
extensive enough. We further derive variational principles for the
axisymmetric modes showing that the unstable modes do occur in discs
with finite albeit small thickness. We give a simple physical picture
of the instability showing how it originates as an unstable
interaction between the central dipole and current loops in the
disc. In section~\ref{sec:equilibria} we describe the specific
equilibrium disc models that we consider and our numerical
calculations. They give spectra of axisymmetric and non-axisymmetric
unstable modes confined to the disc inner regions. These results are
presented in section~\ref{sec:results}. Finally in
section~\ref{sec:discussion} we discuss possible consequences of our
results.

\section{Equilibrium Discs}

\label{sec:equil}

We consider thin disc configurations with an axisymmetric poloidal
magnetic field ${\bmath B}= (B_r,0,B_z).$ Here we use cylindrical polar
coordinates $(r,\phi, z)$. The field is described by a flux function
$\psi$ such that

\be 
B_r= {-1 \over r}{\upartial \psi \over \upartial z}\
{\rm and}\ B_z = {1\over r}{\upartial\psi\over \upartial r}.
\label{psi}
\ee 

\no For this field the current density ${\bmath j}=(0,j_{\varphi},0).$
For an infinitesimally thin disc it is convenient to work with the
vertically integrated azimuthal component of the current density $J,$
where 

\bde 
J=\int^{\infty}_{-\infty} j_{\varphi}{\rm d}z.
\ede 

\no By integrating the azimuthal component of Amp\`ere's law through
the disc, we obtain 

\beq
B_r^{+}= J/2,
\label{br}
\eeq 

\no where $B_r^{+}$ denotes the radial component of the magnetic field
on the upper surface of the disc. Here we have set the magnetic
permeability, $\umu_0,$ to unity. We can recover the equations in MKSA
units by replacing $B$ by $B/\sqrt{\umu_0}$ and $J$ by
$J/\sqrt{\umu_0}$. $B_r$ is antisymmetric with respect to reflection in
the disc mid--plane so that its value on the lower surface of the disc
is $B_r^{-}=-B_r^{+}.$ Thus $B_r$ changes significantly on passing
through the disc in contrast to $B_z$ which, as implied by the
condition $\bnab \cdot {\bmath B}=0,$ changes negligibly on passing
through an infinitesimally thin disc.

\subsection{Force balance}

\label{sec:forceb}

The vertical integration of the radial component of the momentum
equation yields the condition for radial equilibrium as

\be 
\Sigma {\upartial \Phi \over \upartial r}=\Sigma r \Omega^2 + J B_z,
\label{EQ}
\ee 

\no where $\Sigma$ is the surface density, $\Omega$ is the disc
angular velocity, and $\Phi$ is the gravitational potential.

\noindent The above formalism neglects the radial pressure force. An
infinitesimally thin disc can in practice be considered to be
significantly thinner than $c_{\rm s}/\Omega$ ($c_{\rm s}$ being the 
sound speed).
The vertical equilibrium then implies that the mid--plane pressure $P
\sim \left( B_r^{+} \right)^2$, since the magnetic squeezing of the
disc overwhelms its tidal confinement. Thin disc equilibrium
configurations of strongly magnetized discs with the above properties,
 which take account of the vertical structure, have been constructed by 
Ogilvie~(1997). In order to neglect pressure
forces in the radial direction, we require, assuming $B_r^{+} \sim
B_z,$ that $H \ll L_r,$ where $H$ is the semi-thickness of the disc,
and $L_r$ is the scale length of variation in the radial
direction. Thus the approximation scheme becomes better for thinner
discs. We consider equilibria for which the gravity is due to a
central point mass, $M$, such that

\bde 
\Phi =-{GM \over \sqrt{r^2+z^2}}.
\ede 

\no Equilibrium models may be constructed with the surface density,
$\Sigma,$ and integrated current density, $J,$ being specified as
arbitrary functions of $r.$ The flux function with no external sources
is then given by (see Lubow et al.~1994)

\be \psi \left( r,z \right) ={r \over 4 \upi}\int^{R_{\rm o}}_{R_{\rm i}} \int^{2
\pi}_0 {J(r') \cos(\varphi') r' {\rm d}\varphi'{\rm d}r' \over
\sqrt{r^2+r'^2-2rr'\cos(\varphi')+z^2}},
\label{flux}
\ee 

\no where the disc is presumed to have inner and outer boundary radii
$R_{\rm i}$ and $R_{\rm o}$ respectively, the latter possibly being infinite. To
the above internally produced flux, we may add a contribution due to
external sources, $\psi_{\rm {ext}}.$ When the external source is a dipole
at the origin, 

\be 
\psi_{\rm {ext}}=-B_z^{\rm {ext}}(R_{\rm i}){{R_{\rm i}}^3 r^2 \over \left( r^2+z^2 \right)^{3/2}},
\label{extf}
\ee 

\no where $B_z^{\rm {ext}}(R_{\rm i})$ is the external vertical field at the disc inner
boundary. 

\no In the presence of a central dipole, some of the field lines
which cross the inner regions of the disc may join to the dipole in
the centre. Further out, field lines may be open in the case of an
infinite disc or close before the outer boundary when the disc is
finite. Field lines with these properties appropriate to an
equilibrium configuration are illustrated in Fig.~\ref{figf}
below. The sign of the azimuthal current density in this and all other
configurations we consider here is such that the sign of the vertical
field in the inner regions of the disc does not change. That is there
is no X point. This is the situation naturally expected if dipole
field lines diffuse into the disc.

\no For physical consistency, field lines joining the central dipole
should be in a state of isorotation at constant angular velocity, so
$\Sigma$ should be specified accordingly (see
section~\ref{sec:surden}). The condition of isorotation means that the
magnetic field must make an increasingly important contribution to
support the fluid against gravity as $r$ decreases. Spruit~\& Taam~(1990) have
argued that material moving inwards from the outer disc due to angular
momentum transport processes occurring in accretion discs (see
Papaloizou~\& Lin~1995 for a review) can migrate inwards into the
isorotating region due to the action of interchange
instabilities. This may produce a magnetically dominated isorotating
thin disc if the material remains cool. In fact in order to establish
an inner corotating part of the disc, magnetic support against gravity
should not be large at the outer corotation radius. Assuming
$B_r^{+} \sim B_z,$ this requirement gives at that 
radius 

\[ B_z^2 \ll {\Sigma G M\over r^2}.\]

\no Using ${\dot M}=2\upi r v_r \Sigma,$ $v_r$ being the radial
velocity, and the viscous inflow rate $v_r= -\nu /r,$
$\nu$ being the kinematic viscosity, we obtain

\[ B_z^2 \ll {G M |{\dot M}|\over 2\upi\nu r^2}.\]

\no Thus, as indicated above, for fixed stellar properties and disc viscosity,
establishment of an inner corotating region
is favoured at large accretion rates, $|{\dot M}|,$  
and accordingly large disc masses.

\no Open field lines in the outer disc may in principle rotate at any
angular velocity. Field lines that close in the outer regions of a
finite disc may be opened if there are additional external currents
which could be produced by, for example, a wind.

\section{Perturbation Equations}

\label{sec:pert}

We consider linear perturbations of the equilibrium configurations
with a Lagrangian displacement $\bxi$ which, for a razor--thin disc,
has the form

\bde
\bxi = (0,0,\xi_z).
\ede

\no The only non-negligible component is the vertical one which is
independent of $z.$ This displacement belongs to a class such that
$\xi_z$ is even, while the other components are odd with respect to
reflection in the disc mid--plane. These are thus bending modes. We may
also assume that the $\varphi$-dependence of the perturbations is
through a factor $\exp({\rm i}m\varphi)$, $m$ denoting the azimuthal mode
number. From now on this factor will be taken as read and will be
dropped from the perturbations. The Eulerian perturbations of the
various quantities are denoted by a prime.

\no The perturbation of the magnetic field interior to the disc, ${\bmath
B}'$, is related to $\bxi$ by the integration of the linearized
induction equation with respect to time

\be
{\bmath B}' = (B'_r, B'_{\varphi}, B'_z) = \bnab \btimes (\bxi \btimes
{\bmath B})
\ee

\no The non-zero components of ${\bmath B}'$ take the form

\begin{equation}
B'_r = -\xi_z {\upartial B_r\over \upartial z},\ {\rm and }\  
B'_z= {1 \over r}{\upartial (rB_r \xi_z)\over \upartial r}
\label{bpert}
\end{equation}

\no where $B'_z$ is antisymmetric with respect to reflection in
the disc mid--plane and $B'_r$ is symmetric.

\no The vertical component of the perturbed field in the disc must be
matched to the vertical component of the perturbed vacuum field
exterior to the disc. The perturbed vacuum field may be taken to be a
potential field. This is the case even when the disc takes the form of
an annulus making the vacuum multiply connected, because the symmetry
properties of the magnetic field perturbation make it
circulation--free. On the upper disc surface we therefore have $ \upartial
\Phi'_{\rm M}/\upartial z = B'^{+}_z$, where $B'^{+}_z$ is the value of the
vertical field perturbation just outside the disc surface and
$\Phi'_{\rm M}$ is the magnetic potential associated with the external field
perturbation. Continuity of the vertical field component at the upper
disc surface implies that $\Phi'_{\rm M}$ can be found using
(\ref{bpert}). Thus we obtain that on the upper surface

\be
\frac{\upartial \Phi'_{\rm M}}{\upartial z} = {1 \over r}{\upartial (rB^{+}_r
\xi_z)\over
 \upartial r}.
\label{phim}
\ee

\no The corresponding equation with $+\rightarrow -$ applies on the
lower surface. Finding $\Phi'_{\rm M}$ is entirely analogous to finding the
gravitational potential $\Psi$ due to a disc surface density
distribution (see Tagger et al.~1990, Spruit et al.~1995). If
$\Phi'_{\rm M}$ is taken to be equivalent to $\Psi,$ the appropriate surface
density is equivalent to $B'^{+}_z/(2 \pi G),$ where $G$ is the
gravitational constant. Thus $\Phi_{\rm M}'$ may be written in the form of
a Poisson integral

\begin{equation} 
\Phi'_{\rm M} = -{1 \over 2\pi} \int^{R_{\rm o}}_{R_{\rm i}}\int^{2\pi}_0
{B'^{+}_z(r') \cos(m\varphi') r' {\rm d}r' {\rm d}\varphi' \over 
\sqrt{r'^2+r^2-2rr'\cos(\varphi')+z^2}} .
\label{mpot}
\end{equation}

\no The radial component of the magnetic field perturbation on either
the upper or lower surfaces of the disc, just outside the disc, is
then given by

\be
B'^{+}_r= B'^{-}_r =\left({\upartial \Phi'_{\rm M} \over \upartial
r}\right)_{z=0}.
\label{pot}
\ee

\subsection{Vertical component of the equation of motion}

In general, the vertical component of the Lorentz force per unit volume
is

\be 
F_z = -{1 \over 2}{\upartial B_r^2 \over \upartial z}+ B_r
{\upartial B_z \over \upartial r}.
\ee 

\no Perturbing and integrating this vertically through the disc gives

\be 
\int^{\infty}_{-\infty} F'_z {\rm d}z = -2 B_r^{+} B_r'^{+}- 2 \xi_z B_r^{+}
{\upartial B_z \over \upartial r}
\label{force}
\ee 

\no where we have assumed that $B_z$ is almost independent of $z$ in the
disc. The perturbed vertically integrated $z$-component of the equation
of motion is 

\be 
\Sigma{{\rm D}^2 \xi_z \over {\rm D}t^2}= -\Sigma \left({\upartial^2
\Phi \over \upartial z^2}\right)_{z=0} \xi_z +
\int^{\infty}_{-\infty} F'_z {\rm d}z
\label{mot}
\ee 

\no where ${\rm D}$ denotes the convective derivative.

\no The coefficients of equation~(\ref{mot}) are independent of $t$.
We can therefore look for solutions in the form of normal modes. In
this case the time--dependence of the perturbed quantities is taken to
be through a factor $\exp({\rm i}\sigma t)$, where $\sigma$ is the
eigenfrequency of the mode. Using this together with the fact that for
a point mass potential

\bde
\left({\upartial^2 \Phi \over \upartial
z^2}\right)_{z=0}={GM/r^3}=\Omega_{\rm K}^2,
\ede

\no where $\Omega_{\rm K}$ is the keplerian angular velocity,
equation~(\ref{mot}) becomes the normal mode equation

\beq
\left[(\sigma+m\Omega)^2-\Omega_{\rm K}^2- {2B_r^{+}\over \Sigma}
{\upartial B_z\over \upartial r}\right]\xi_z\!=\!{2B_r^{+}\over
\Sigma}\left({\upartial\Phi'_{\rm M}\over\upartial r}\right)_{z=0}\!\!\!\!\!\!\!.
\label{mod}
\eeq

\no Equation~(\ref{mod}) together with~(\ref{mpot}) and~(\ref{bpert})
constitutes a linear eigenvalue problem with $\sigma$ as the eigenvalue
and $\xi_z$ as the eigenfunction.

\subsection{Local Dispersion Relation}

\label{sec:dispersion}

We can derive a local dispersion relation from~(\ref{mod}) by adopting
perturbations of the form $\xi_z \propto \exp({\rm i}kr),$ where $k$ is the
radial wavenumber, assumed to be $\gg |m|/r.$ We comment that because
the evaluation of $\Phi'_{\rm M}$ is equivalent to calculation of the
gravitational potential due to a surface density $B'^{+}_z/(2\upi G),$
the situation here is closely analogous to that for bending modes in a
self-gravitating disk (see Hunter~\& Toomre~1969, Shu~1984). The
integral in~(\ref{mpot}) can thus be calculated using the WKB
approximation to give

\bde \Phi'_{\rm M} = -B'_z/|k| = -{\rm i}kB_r^{+}\xi_z/|k|  \ede

\no where $\Phi'_{\rm M}$ is now the amplitude of the mode with radial
wavenumber $k$. The local dispersion relation derived from~(\ref{mod})
is then

\be (\sigma+m\Omega)^2= \Omega_{\rm K}^2 +{2B_r^{+}\over \Sigma} {\upartial
B_z\over \upartial r} + {2(B_r^{+})^2\over \Sigma}|k|.
\label{disp} \ee

\no Instability ensues on the existence of at least one mode with
growth rate ${\rm i}(\sigma+m\Omega) >0$, for which we require
$(\sigma+m\Omega)^2 <0$.  As the last term on the right-hand side
of~(\ref{disp}) is positive definite and therefore stabilising, the
condition for instability becomes

\be 
\Omega_{\rm K}^2 +{2B_r^{+}\over \Sigma} {\upartial B_z\over \upartial r}
< 0. 
\label{cond}
\ee 

\no When the gravity is due to a central point mass,~(\ref{EQ}) gives

\be 
r(\Omega_{\rm K}^2-\Omega^2)= {2B_r^{+}B_z\over \Sigma},
\label{fb}
\ee 

\no so that~(\ref{cond}) may be expressed in the equivalent form 

\be 
{\Omega_{\rm K}^2\over(\Omega_{\rm K}^2-\Omega^2)} +{r\over B_z} {\upartial
B_z\over \upartial r} < 0. 
\label{cond1}
\ee 

\no We comment that in the non-rotating case ($\Omega=0$),
(\ref{cond1})~becomes the same condition as that given by Wu~(1987)
and Lepeltier~\& Aly~(1996) who considered non-rotating current
sheets. The condition~(\ref{cond}) becomes that given by Anzer~(1969) and
Spruit~\& Taam~(1990) provided one sets $\Omega_{\rm K}=0.$ 
 This is because these
authors omitted gravitational restoring forces in the direction
perpendicular to the current sheet.

\no The local criterion for stability becomes satisfied when the disc
is being primarily supported by an external field. For an external
dipole

\bde
{r\over B_z}{\upartial B_z\over \upartial r}= -3.
\ede

\no Then~(\ref{cond1}) will be satisfied when the magnetic field
provides enough support against gravity so that $\Omega^2_{\rm K} >
3\Omega^2/2.$ The latter is satisfied in the interior regions of the
disc where the field lines link to the central dipole and $\Omega$ is
constant with $\Omega_{\rm K}$ increasing inwards.

\no We note that the condition above has been obtained for small
$|k|,$ which is out of the domain of validity of a local
approximation. However, the term in~(\ref{disp}) involving $|k|$ is
proportional to $(B_r^{+})^2$ and it therefore becomes of
decreasing importance for a uniformly rotating magnetically supported
region as this extends inwards towards regions of large $B_z,$ with
the result that instability must eventually ensue. However, the
precise details of onset require explicit calculation.

\subsection {Axisymmetric Modes}

Rigorous global criteria may be obtained in the case of axisymmetric
modes through the existence of variational principles. In this case,
for $m=0,$~(\ref{mod}) can be written in the form

\be 
\sigma^2\xi_z = {\mathcal O}(\xi_z),
\label{modop} 
\ee

\no where the operator ${\mathcal O}$ is self-adjoint in that for
arbitrary eigenfunctions $\xi_z$ and $\eta_z$ the following equality
is satisfied

\bde 
\int^{R_{\rm o}}_{R_{\rm i}}\Sigma r \eta_z^{*}{\mathcal O}(\xi_z){\rm d}r=
\left(\int^{R_{\rm o}}_{R_{\rm i}}\Sigma r \xi_z^{*}{\mathcal
O}(\eta_z){\rm d}r\right)^{*}.
\ede 

\no In this case, a sufficient condition for instability is that, for
any $\xi_z,$

\begin{eqnarray}
\int^{R_{\rm o}}_{R_{\rm i}} \Sigma r \xi_z^{*} {\mathcal O}(\xi_z)
{\rm d}r & = & 
\int^{R_{\rm o}}_{R_{\rm i}} \Sigma r \left( \Omega_{\rm K}^2+ {2B_r^{+} \over \Sigma}
{\upartial B_z \over \upartial r} \right) |\xi_z|^2 {\rm d}r
\nonumber \\
& + & {1 \over \pi} \int^{R_{\rm o}}_{R_{\rm i}} \int^{R_{\rm
o}}_{R_{\rm i}} \int^{2\pi}_0 {
\left( B'^{+}_z(r) \right)^{*} B'^{+}_z(r') \cos(m\varphi') rr' {\rm
d}r{\rm d}r'
{\rm d}\varphi' \over \sqrt{r'^2+r^2-2rr'\cos(\varphi')}} <0 .
\label{var}
\end{eqnarray} 

\no On insertion of suitable local trial functions, this gives the
same condition as~(\ref{disp}).

\subsection{Relation to thick disc analysis}

We here note that the above variational principle may also be derived
from the general variational principle of Papaloizou~\&
Szuszkiewicz~(1992) for stability to adiabatic perturbations of a
general differentially rotating equilibrium with a purely poloidal
magnetic field, when the thin disc limit is taken.

\no This establishes that the results are not an artefact of the use of
the razor--thin disc approximation and vertically averaged equations
from the outset. A sufficient condition for stability to axisymmetric
modes is that for any trial $\bxi$:

\begin{equation}
\int \xist \cdot {\bmath L}(\bxi) {\rm d}V < 0 
\label{PCOND}
\end{equation} 

\no where ${\bmath L}$ is a linear operator, defined in Papaloizou~\&
Szuszkiewicz~(1992) and below, similar in principle to ${\mathcal O}$ but
which depends on both $r$ and $z$. The integral is now a volume
integral (see below).

\no The condition~(\ref{PCOND}) is also necessary for stability to
density perturbations with odd symmetry with respect to reflection in
the equatorial plane such that $(\xi_r,\xi_{\varphi},\xi_z)\rightarrow
(-\xi_r,-\xi_{\varphi},\xi_z).$ This corresponds to modes of the
symmetry type considered in this paper. For these we have instability
if

\begin{equation}
-\int \xist \cdot {\bmath L}(\bxi) {\rm d}V = \int \left( 
\frac{|P'|^2}{\Gamma_1 P} + \rho Q(\bxi,\xist) - 
\frac{j_{\varphi }}{r} \left( \bxi \cdot \bnab(\psi')^* + 
\xist \cdot \bnab \psi' \right) + \frac{|\bnab \psi'|^2}{ r^2}
\right) {\rm d}V <0 .
\label{LIS}
\end{equation} 

\no Here, $\rho, \Gamma_1$ and $\psi'$ denote the density, specific
heat ratio and magnetic flux perturbation respectively. For a fluid
with smoothly vanishing density and pressure at the boundary all
integrals, other than the last, are taken over the fluid volume $V$. The
last integral which is related to the magnetic energy associated with
the perturbation must be taken over the whole of space excluding any
perfect conductors. The pressure perturbation is given by

\bde 
P' =-\Gamma_1 P \bnab \cdot \bxi - \bxi \cdot \bnab P.  
\ede

\noindent The quadratic form $Q(\bxi,\xist)$ is given by

\begin{equation}
Q(\bxi,\xist) = r(\xist \cdot \bnab r)(\bxi \cdot \bnab \Omega ^2) -
\xist \cdot \frac{\bnab P}{\rho} \bxi \cdot 
\left( \frac{\bnab P}{\Gamma _1 P}-\frac{\bnab \rho }{\rho } \right) 
-(\psi')^* \bxi \cdot \bnab \left(\frac {j_{\varphi }}{r\rho }\right).
\label{QQ} 
\end{equation}

\no In our case we adopt $\bxi = (0,0,\xi_z)$ with $\xi_z$ being real and
depending only on $r.$ For this trial function, use of $\psi' = -\bxi \cdot
\bnab \psi$ inside the ideal fluid, together with vertical hydrostatic
equilibrium, gives 

\be 
-\int \xist \cdot {\bmath L}(\bxi) {\rm d}V = \int \left( \rho 
{\upartial^2 \Phi \over \upartial z^2} + {\upartial B_z \over \upartial r} 
{\upartial B_r \over \upartial z} \right) \xi_z^2 {\rm d}V + 
\int {1 \over r^2} 
\left( {\upartial \over \upartial r}(\xi_z r B_r) \right)^2 dV +
\int_{\rm {vac}} {\bmath B}'^2 {\rm d}V. 
\label{var1}
\ee 

\no Here all integrals except the last are taken over the fluid disc
while the last is taken over the vacuum outside.

\noindent We see that there is a close correspondence
between~(\ref{var1}) and~(\ref{var}) which applies to the razor--thin
disc. Each of the integrals in~(\ref{var1}) apart from the second,
which becomes vanishingly small compared to the first in the thin disc
limit, approaches the corresponding integral in~(\ref{var}). Thus the
instabilities we find are not due to the {\em ab initio} assumption of a
razor--thin disc.

\subsection{A simple picture of the instability}

Here we show how the condition for instability~(\ref{cond}) can be
found from a simple argument relating to the interaction of a current
loop with the central dipole. For the dipole field interior to the
loop to be of the same orientation as the field produced by the loop
itself (no X--point), the dipole moment must have the opposite sign to
the azimuthal current density which produces an unstable
interaction. A current loop at $(r,z)$ with total current $I$
produces a magnetic field at the origin with a vertical component

\be 
B_0= {I r^2 \over 2(z^2+r^2)^{3/2}}.
\ee

\no The energy of an anti-aligned dipole of dipole moment $\mu$ situated 
at the centre of the loop is

\bde
W=\mu B_0= {\mu I r^2 \over 2(z^2+r^2)^{3/2}}.
\ede

\no If we suppose that the loop lies initially at $z=0,$ and that is
then displaced vertically up to $z,$ where $z$ is small compared to
$r$, the change on the energy is

\bde
\delta W = -{3\mu I z^2\over 4r^3}.
\ede

\no To obtain the total change in energy we must add the change in
gravitational potential energy $m_{\rm s}\Omega_{\rm K}^2 z^2/2,$
where $m_{\rm s}$ is
the loop mass. Writing $m_{\rm s}=2\upi\Sigma r{\rm d}r,$ and $I=
J{\rm d}r,$ we obtain
for the total energy change

\bde 
\delta W = \left(-{3 \mu J \over 4\upi \Sigma r^4}+\Omega_K^2
\right)\upi\Sigma rz^2{\rm d}r.
\ede

\no Using~(\ref{br}) and the expression for the external vertical
field produced by the dipole, $B_z^{\rm {ext}}= \mu/(4\upi r^3),$ we obtain

\bde 
\delta W = \left( {2 B_r^{+}\over \Sigma}{\upartial B_z^{{\rm ext}}\over
\upartial r} +\Omega_K^2 \right)\upi\Sigma rz^2{\rm d}r.
\ede 

\no The natural condition for instability is that energy be released
on making the displacement, or in our case $\delta W < 0.$ This gives
the same condition as~(\ref{cond}) if we replace $B_z$ with the
external vertical field which is the one that dominates under
conditions of strong magnetic support.

\noindent Thus we see that the generic instability can be understood
in terms of an unstable interaction between the disc current and the
central dipole. In the simple example described above, when the system
passes through marginal stability because of say an increasing dipole
moment, a bifurcation occurs such that there is a new stable
equilibrium for the current loop lying off the original mid--plane. This
suggests that in the case of a continuous disc, a warped structure may
be taken--up in which the inner regions are elevated above or below
the symmetry plane, facilitating eventual motion towards the central
object along field lines. We now go on to discuss numerical
calculations of normal modes for some specific models.

\section{ Numerical Calculations}

\label{sec:equilibria}

\subsection{Equilibrium disc current and flux function}

In order to solve~(\ref {mod}) we need to specify functional forms for
$\Sigma$ and $J$. The radial and vertical components of the magnetic
field are then determined from $\psi$ which is calculated
using~(\ref{flux}). To evaluate the integral in the right hand side
of~(\ref {flux}) we use the method outlined in Lubow et
al.~(1994). The region between $R_{\rm i}$ and $R_{\rm o}$ is divided into $N_r$
equally spaced grid points with separation ${\rm d}r.$ The surface current
density $J$ is discretised in the form of concentric ring currents
$I(r_i) = J(r_i){\rm d}r$ where subscripts $i$ (and $j$ later) denote values
calculated at the $i^{\rm {th}}$ ($j^{\rm {th}}$) grid
point. Equation~(\ref{flux}) is then approximated by

\beq
{{\psi}}(r_i) ={1 \over{4\upi}} r_i \sum_{j}^{}K_{ij}I(r_j)r_j
\label{SUM}
\eeq

\noindent where $K_{ij}$ is the discretised form of the integral with
respect to $\varphi$ in~(\ref{flux}); this can be written in terms of
elliptic integrals (see Jackson~1975). We thus write

\bde 
K_{ij} = \frac{\displaystyle 4}{\displaystyle
{\sqrt{{r_i}^2+{r_j}^2+2r_i\ r_j}}} \frac{\displaystyle\left[
{(2-k^2)E_1(k^2)-2E_2(k^2)}\right]}{\displaystyle {k^2}} 
\ede

\noindent where $E_1(k^2)$ and $E_2(k^2)$ are elliptic integrals of
the first and second kind respectively, and

\bde 
k^2 = \frac{\displaystyle {4r_i\ r_j}}{\displaystyle
{{r_i}^2+{r_j}^2+2r_ir_j}} 
\ede

\noindent In the numerical calculations performed here, the surface
current density $J$ is, to within a scaling factor, taken to be given
by

\bde
{J} = f_{\rm J}(r_i) - f_{J}(R_{\rm o}+{\rm d}r)
\ede

\noindent where 
 
\beq 
f_{\rm J}(r_i) = {c_1 \sqrt{4\upi}}\ [{\exp}({c_2 \cdot nc(r_i -
r_{\rm {mid}})^2/{R_{\rm o}}^2}) + {(r_i/R_{\rm o})}^{c3 \cdot nc}]^{-1/nc}. 
\label{JPROF} 
\eeq

\noindent Both $J$ and $\Sigma$ can be taken to vanish at some point
simultaneously in the same manner so as to retain a finite Lorentz
force. We have chosen this point to be at a fictitious additional grid
point at $R=R_{\rm o}+{\rm d}r$ for numerical convenience.

\noindent The values of the constants $c_1\ldots c_3$, $r_{\rm {mid}}$ and
$nc$ are :

\(\begin{array}{lll}
c_1 = 10    & c_2 = 0.1   & c_3  = 2  \\
nc  = 4     & r_{\rm {mid}} = 0.9 R_{\rm o} &
\end{array}\)

\noindent The radial distribution of ${\psi}$ is calculated from
equation~(\ref{SUM}). The corresponding field lines in the $z^+ $
quarter of the disc are the contours of ${\psi}(r,z) = const$. To the
above internally produced magnetic flux we add a contribution from a
dipole at the origin given by~(\ref{extf}). Since the disc is thin, we
shall neglect the radial magnetic field produced by this dipole and
consider only the vertical component. 

\subsection{Dimensionless variables}

We define a modified epicyclic frequency $\kappa_{\rm m}$ such that

\be \kappa^2_{\rm m} = \Omega_{\rm K}^2 + \frac{2 B_r^+}{\Sigma} \frac{\upartial
B_z}{\upartial r} 
\label{kappa} \ee

\no and consider the dimensionless variables $x=r/R_{\rm o},$
$\bar{\kappa}=\kappa_{\rm m}/\Omega_0,$ $\bar{\sigma}=\sigma/\Omega_0,$
$\bar{\Omega}=\Omega/\Omega_0$, $\bar{B}_r^+=B_r^+/B_0$,
$\bar{\psi}=\psi/(B_0 {R_{\rm o}}^2)$ and $\bar{\Sigma}=\Sigma/\Sigma_0$,
with $\Omega_0=\Omega_{\rm K}(R_{\rm o})$ and $B_0$ and $\Sigma_0$ being some
fiducial values of the magnetic field and surface mass density
respectively. The normal mode equation~(\ref{mod}) can then be written
in the dimensionless form

\be \left( \bar{\kappa}^2 - \left( \bar{\sigma}+m \bar{\Omega}
\right)^2 \right) \xi_z = - \beta_0 \frac{2 \bar{B}_r^+}{\bar{\Sigma}}
\frac{\upartial \bar{\Phi}'_{\rm M}}{\upartial x} 
\label{modd} \ee

\no were $\bar{\Phi}'_{\rm M}={\Phi}'_{\rm M}/(B_0 R_0)$ and $\beta_0$ is a constant
which measures the relative strength of the magnetic and centrifugal
forces (the larger $\beta_0$, the larger the magnetic support):

\bde
\beta_0 = \frac{B_0^2 R_{\rm o}^2}{GM \Sigma_0}.
\ede

\no The dimensionless external magnetic flux in the disc $(z=0)$ takes
the form $\bar{\psi}_{{\rm ext}} = -\bar{\psi}_0/x$. Equilibrium models
which have scaled specified profiles for disc current and surface
density are parametrised by the two parameters, $\bar{\psi}_0$ which
measures the ratio of dipole to disc fields and $\beta_0$ which can be
thought of scaling the disc surface density to provide a desired
degree of magnetic support.

\begin{figure}
\centerline{\epsfig{file=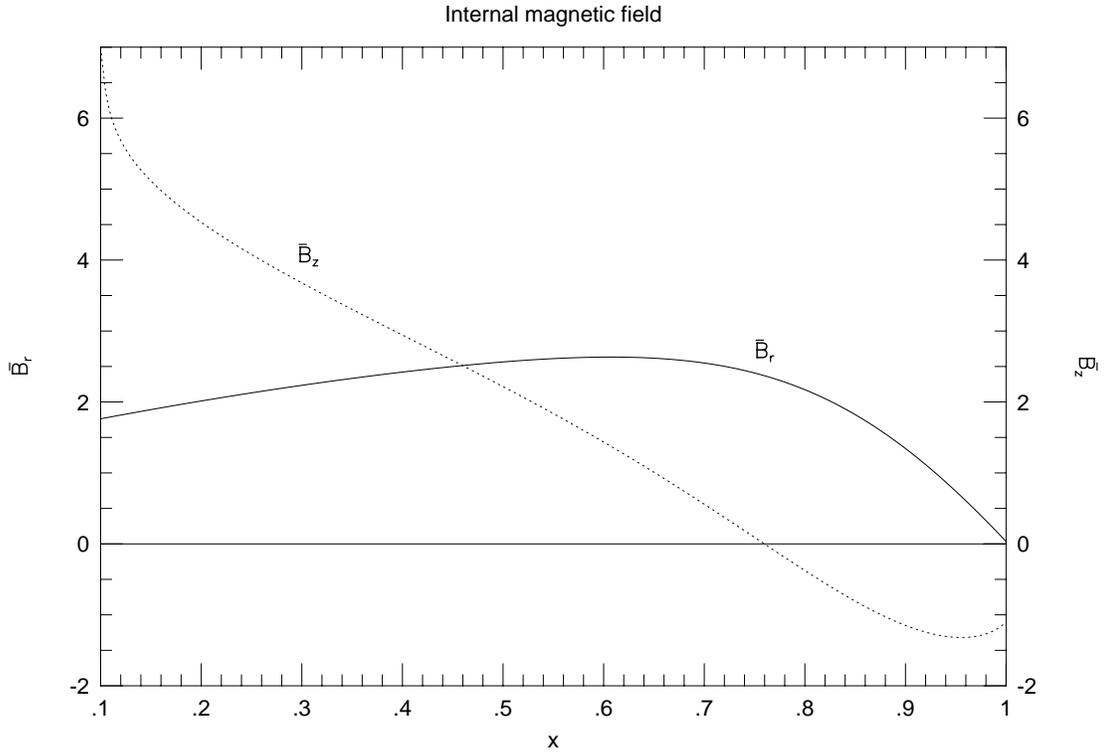,height=120mm,angle=0}}
\caption{Radial (${\bar B}_r$) and vertical (${\bar B}_z$) components
of the magnetic field due to the disc currents. The magnitude of the 
dimensionless dipole field at the centre is $30$ for models of type~I 
and $60$ for models of type~II}
\label{fige}
\end{figure}

\begin{figure}
\noindent
\bmin{.48}
\centering\epsfig{file=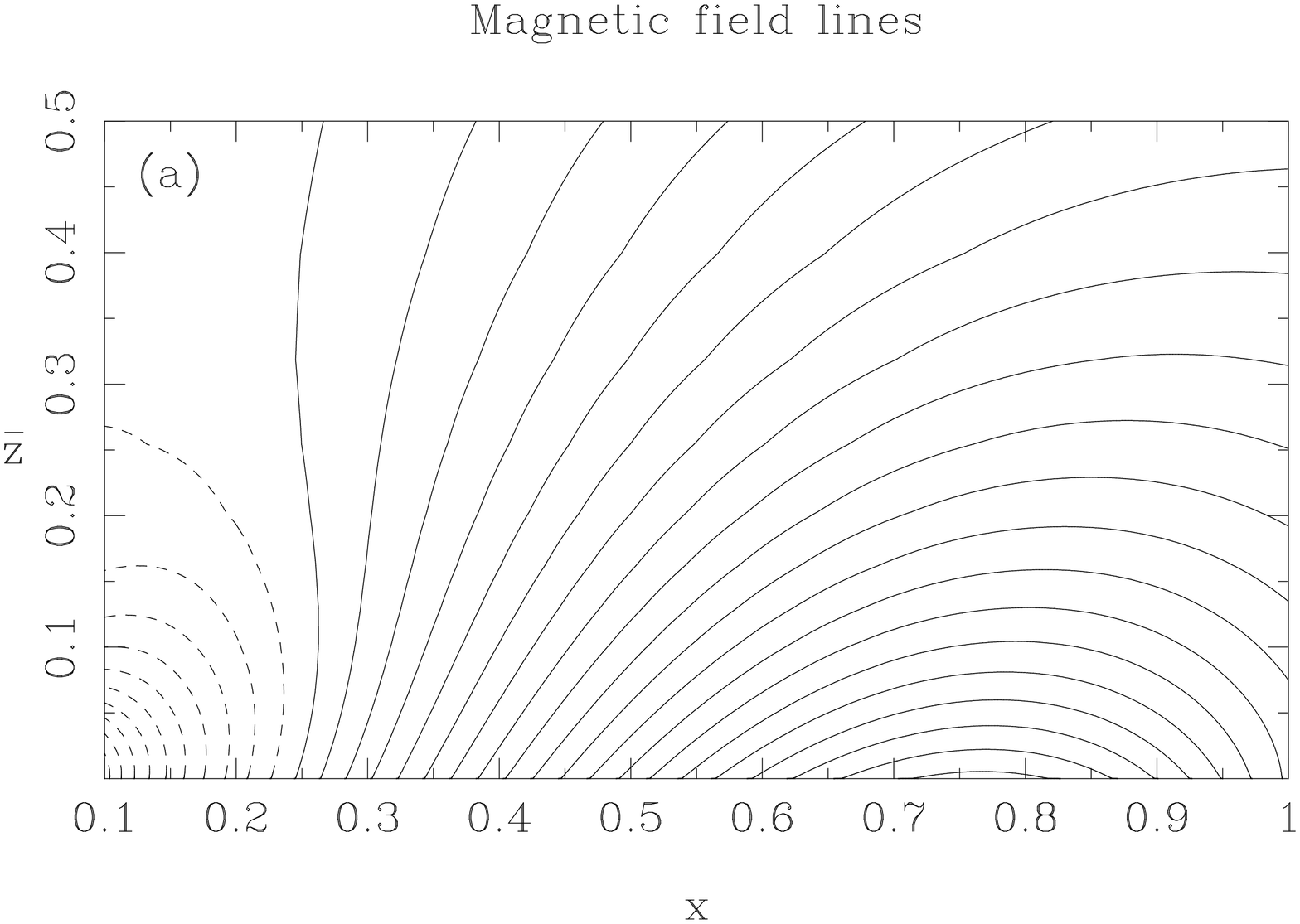,height=75mm,width=\linewidth,angle=270}
\emin \hfill
\bmin{.48}
\centering\epsfig{file=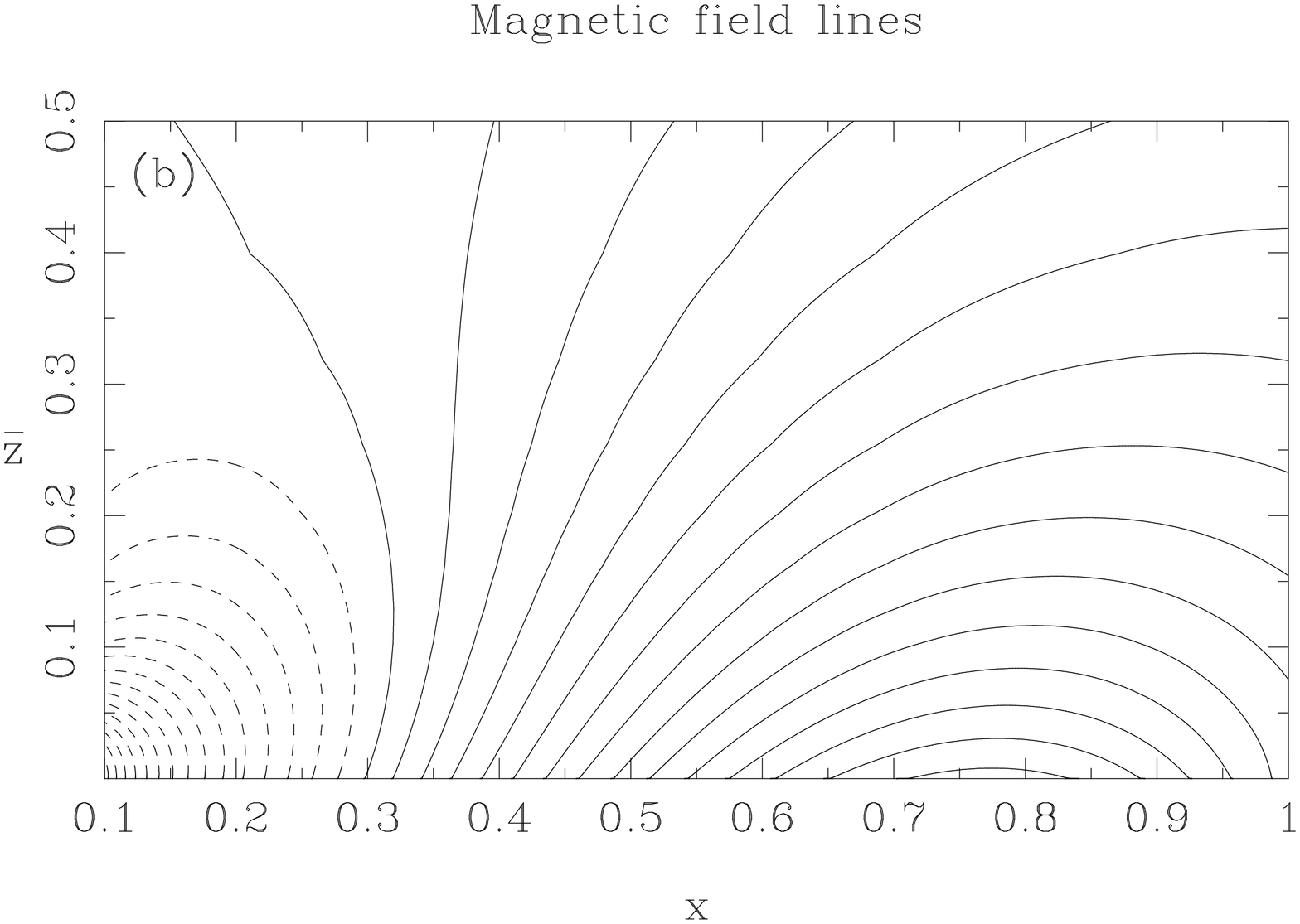,height=75mm,width=\linewidth,angle=270}
\emin
\caption{Magnetic field lines for (a) models of type~I and (b) models of
type~II.}
\label{figf}
\end{figure}

\no In section~\ref{sec:results} we present the results of normal
mode calculations using equilibrium models with two different values
of $\bar{\psi}_0,$ models of type~I with $\bar{\psi}_0=0.03$ and
models of type~II with $\bar{\psi}_0=0.06.$ Various values of
$\beta_0$ have been considered. We plot the dimensionless components
of the magnetic field produced by the disc currents in
Fig.~\ref{fige}. The magnitudes of the dipole field at the innermost
radius are given for models of type~I and~II for comparison. The
contours defined by $\bar{\psi}_{\rm {ext}}+\bar{\psi} = const$ are
plotted in Fig.~\ref{figf}(a) for models of type~I and in
Fig.~\ref{figf}(b) for models of type~II.

\subsection{Surface density and Angular velocity}

\label{sec:surden}

The surface density and angular velocity are calculated in the
following way. The magnetic support, $s_{\rm m},$ is defined as the ratio of
magnetic to centrifugal forces in the condition for radial
equilibrium:

\bde
|s_{\rm m}|=\frac{2 |\bar{B}_r^+ \bar{B}_z| x^2}{\bar{\Sigma}}
\ede

\no where $\bar{B}_z=B_z/B_0$. We first fix the value of
$|s_{\rm m}|.$ It is reasonable to assume that in an accretion disc the
magnetic support due to the internal field is larger in the inner
parts of the disc than in the outer parts. This is indeed the case if
the magnetic field in the disc is advected radially due to the
accretion (Lubow et al.~1994, Reyez-Ruiz~\& Stepinski~1996, Agapitou~\& 
Papaloizou~1996). In
addition, the dipole increases the magnetic support in the central
region of the disc. For these reasons we take $|s_{\rm m}|$ to be a
decreasing function of $x$. $\bar{\Sigma}$ is then calculated from
$|s_{\rm m}|$ and adjusted in order to make sense physically. $\bar{\Omega}$
is then deduced from the dimensionless form of the radial
equilibrium~(\ref{EQ}) 

\be
\bar{\Omega}^2=\bar{\Omega}_{\rm K}^2-\beta_0 \frac{s_{\rm m}}{x^3},
\label{EQd}
\ee 

\no where~(\ref{br}) has also been used. As mentioned above
(section~\ref{sec:forceb}), the inner parts of the disc where the
magnetic field lines are linked to the dipole corotate with the
dipole. The dipole flux is negative whereas the disc internal flux is
positive. In the inner parts of the disc the total flux is then
negative and in the absence of singular points the field lines are
connected to the dipole. In contrast the total flux in the outer parts
of the disc is positive and the field lines are not linked to the
dipole. We then fix the angular velocity $\bar{\Omega}$ to be constant
in the part of the disc where the total flux is negative, equal to its
value at the point where the flux vanishes. Given this new
$\bar{\Omega}$ profile, we then recalculate $\bar{\Sigma}$
using~(\ref{EQd}).

\begin{figure}
\centerline{\epsfig{file=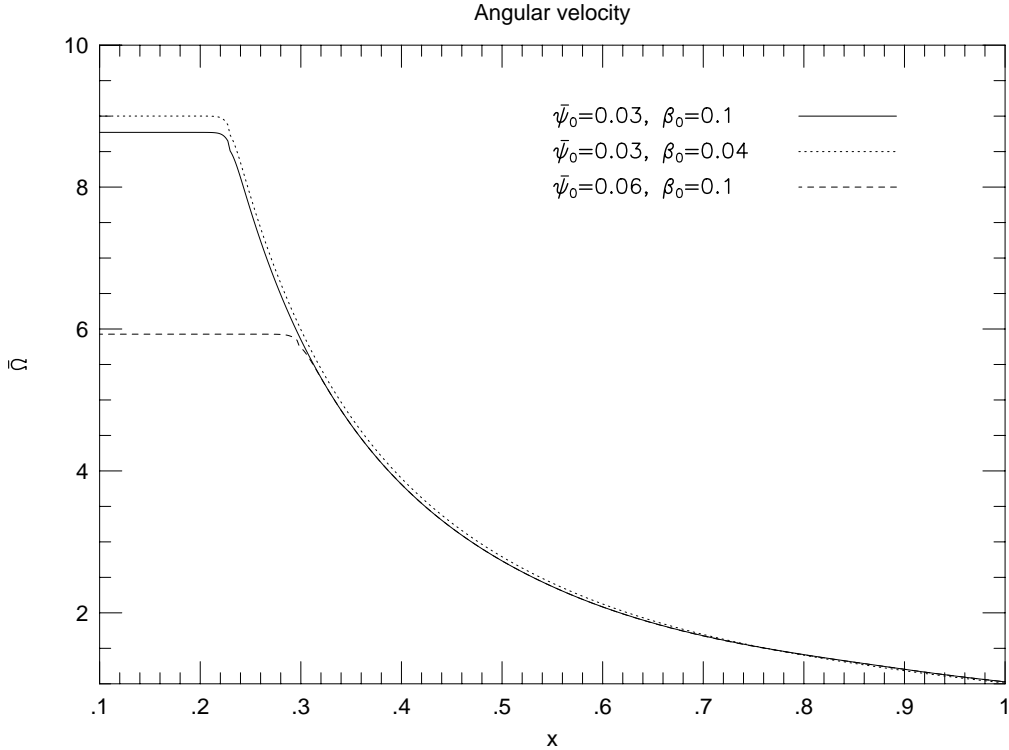,height=120mm,angle=0}}
\caption[]{Angular velocity $\bar{\Omega}$ for $\bar{\psi}_0=0.03$ and
$\beta_0=0.1$ (solid line), $\bar{\psi}_0=0.03$ and $\beta_0=0.04$
(dotted line) and $\bar{\psi_0}=0.06$ and $\beta_0=0.1$ (dashed
line).}
\label{figa}
\end{figure}

\begin{figure}
\centerline{\epsfig{file=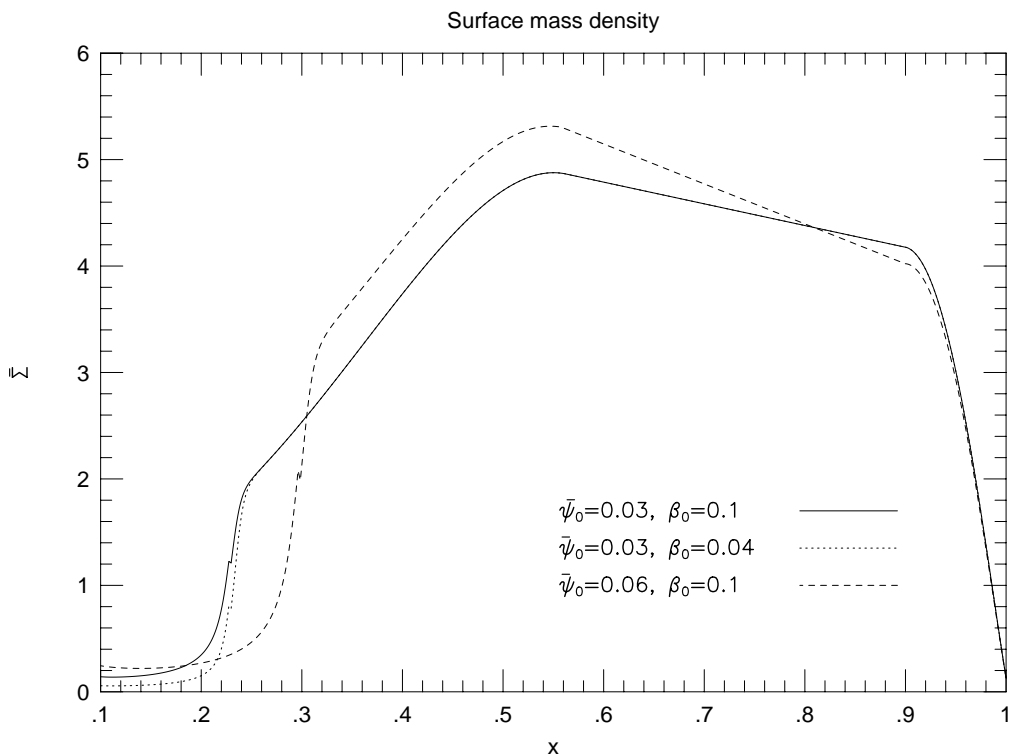,height=120mm,angle=0}}
\caption[]{Surface density $\bar{\Sigma}$ for $\bar{\psi}_0=0.03$ and
$\beta_0=0.1$ (solid line), $\bar{\psi}_0=0.03$ and $\beta_0=0.04$
(dotted line) and $\bar{\psi_0}=0.06$ and $\beta_0=0.1$ (dashed
line).}
\label{figb}
\end{figure}

\begin{figure}
\centerline{\epsfig{file=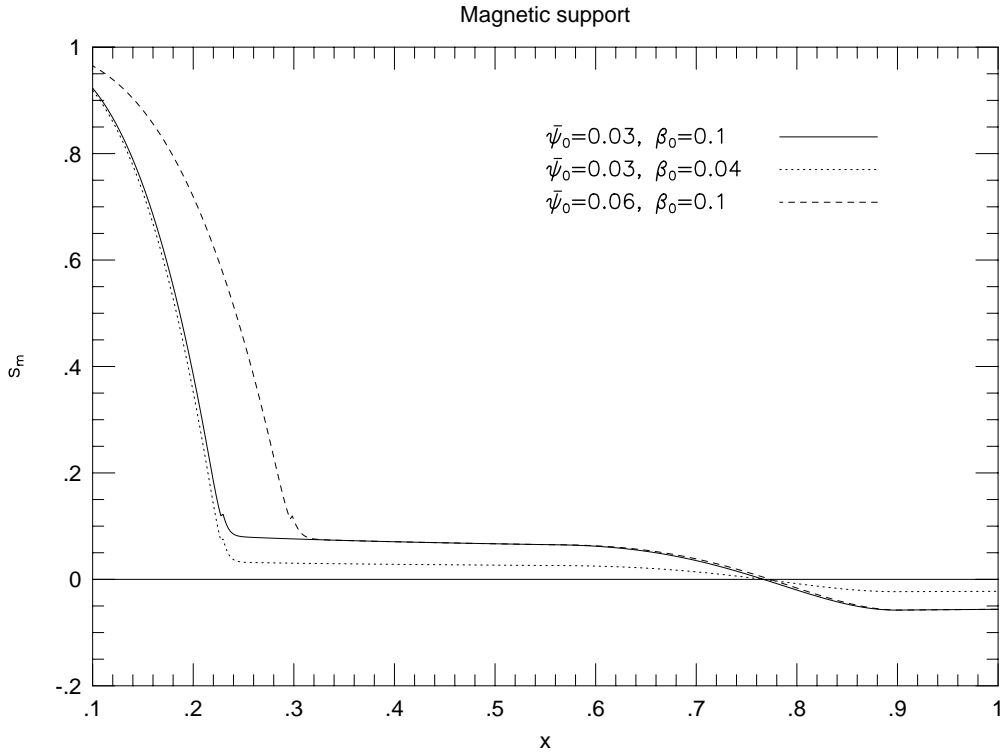,height=120mm,angle=0}}
\caption[]{Magnetic support $s_m$ for $\bar{\psi}_0=0.03$ and
$\beta_0=0.1$ (solid line), $\bar{\psi}_0=0.03$ and $\beta_0=0.04$
(dotted line) and $\bar{\psi_0}=0.06$ and $\beta_0=0.1$ (dashed
line).}
\label{figc}
\end{figure}

\begin{figure}
\centerline{\epsfig{file=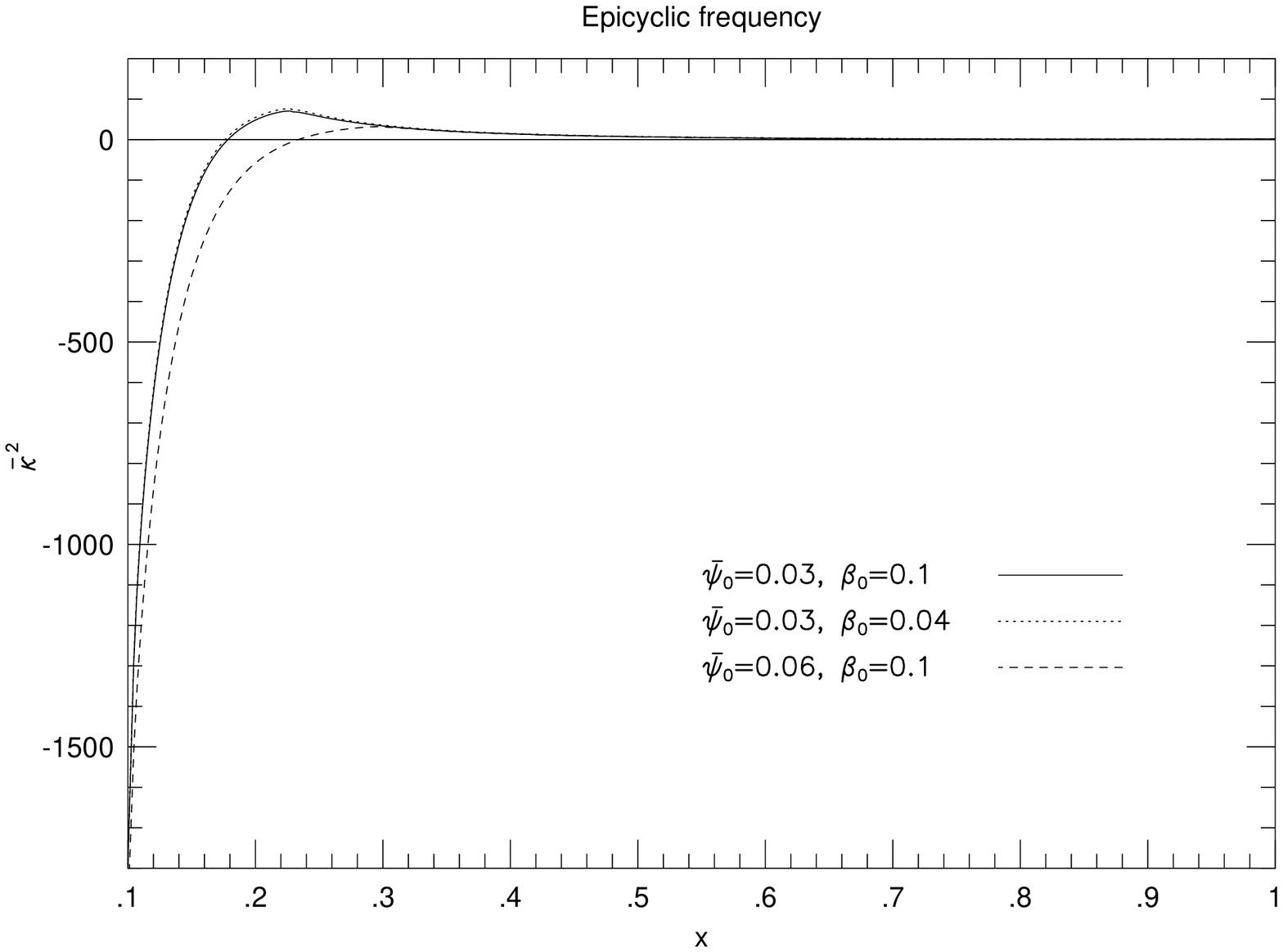,height=120mm,angle=0}}
\caption[]{Square of the epicyclic frequency, $\bar{\kappa}^2,$ for
$\bar{\psi}_0=0.03$ and $\beta_0=0.1$ (solid line),
$\bar{\psi}_0=0.03$ and $\beta_0=0.04$ (dotted line) and
$\bar{\psi_0}=0.06$ and $\beta_0=0.1$ (dashed line).}
\label{figd}
\end{figure}

We have performed calculations for disc models with a magnetic field
configuration of the type described above. For all the models the
inner disc radius $x_{\rm {in}}=0.1$ and the outer disc boundary is at
$x_{\rm o}=1$. Figures~\ref{figa} to~\ref{figd} show respectively 
the equilibrium profiles of
$\bar{\Omega},$ $\bar{\Sigma},$ $s_{\rm m}$ and $\bar{\kappa}^2$ 
used for models of type~I and~II with $\beta_0=0.1$ or $0.04.$

From Fig.~\ref{figc} we see that the inner parts of the disc are
dominated by the dipole field. Indeed $\beta_0$, which controls the
strength of the disc magnetic field, does not affect the magnetic
support close to the inner edge of the disc.

\section{Normal mode calculations}
 
\label{sec:results}

We have performed global normal mode calculations for the disc models
described above. Both axisymmetric modes and modes with small values
of $m$ have been considered and unstable spectra found.

\no We solve equation~(\ref{modd}) by dividing the radial interval
$\left[x_{\rm {in}},x_{\rm o} \right]$ into a grid of $n_r$ points at positions
$\left( x_i \right)_{i=1 \ldots n_r}$ with a spacing ${\Delta x}_i =
x_{i+1}-x_{i}$. The disc boundaries are at $x_1=x_{\rm {in}}=0.1$ and
$x_{n_r}=x_{\rm o}=1$.

\no We approximate the integral in the expression~(\ref{mpot}) for 
$\Phi'_{\rm M}$ by a sum of functional values at the points $\left( x_i
\right)_{i=1 \ldots n_r}$. Since the integrand involves $\xi_z$
through $B'^{+}_z$ (see~(\ref{bpert})), the discretised form of
equation~(\ref{modd}) is a system of $n_r$ equations of the form

\be \left[ \left( \bar{\sigma} + m \bar{\Omega}_i \right)^2 -
\bar{\kappa}^2_i \right] \xi_{z,i} = \sum_{j=1}^{n_r} A_{ij}
\xi_{z,j}, \; \; \; i=1 \ldots n_r \label{system} \ee

\no where the subscript $i$ (or $j$) denotes the value of the function
at the point $x_i$ (or $x_j$) and $\textbfss{A}$ is a matrix which depends
on the characteristics of the disc. If $m=0$, (\ref{system}) is a $n_r
\times n_r$ eigensystem with eigenvalue $\bar{\sigma}$ and
eigenfunctions $\xi_{z,i}, i=1 \ldots n_r$. For $m$ non-zero, we set

\bde u_i=\frac{\bar{\sigma}}{2m \bar{\Omega_i}} \xi_{z,i}. \ede

\no The system~(\ref{system}) is then equivalent to a $2n_r \times
2n_r$ eigensystem with eigenvalue $\bar{\sigma}$ and eigenfunctions
$\xi_{z,i}, i=1 \ldots n_r$ and $u_i, i=1 \ldots n_r$. The
eigenvalues are calculated numerically using the QR algorithm for real
Hessenberg matrices given by Press et al.~(1986). Once $\bar{\sigma}$
is found, the solution of the system~(\ref{system}) gives $\xi_{z,i},
i=1 \ldots n_r$.

\no Table~\ref{table1} summarises the characteristics of the disc
models considered. For a mode to be unstable we require that
${\rm Im}(\bar{\sigma})<0$. The table also gives for each model the
eigenvalues $\bar{\sigma}$ for the unstable modes found. The real part
of $\bar{\sigma}$ is the frequency of the mode and the imaginary part
relates to its growth rate. In the column which contains $n_r$, we
have indicated whether the grid is uniform~(u) or non-uniform~(n).
 
\begin{table}
\caption{Characteristics of the disc models and values of $\bar{\sigma}$ for 
unstable modes.}
\begin{tabular} {cccccccccccccccc} \hline 

Model & $\bar{\psi}_0$ & $\beta_0$ & $m$ & $n_r$ & $\bar{\sigma}$ (the
unresolved modes are followed by *)\\ \hline

I1  & 0.03 & 0.1 & 0   & 799 (u)  & (0,-37.44) (0,-17.31) (0,-7.62) \\ 
I2a & ---  & --- & 1   & 799 (u)  & (-8.77,-37.22) (-8.77,-17.80) (-8.77,-7.31) \\
I2b & ---  & --- & --- & 1175 (n) & (-8.79,-36.90) (-8.79,-17.47) (-8.79,-6.72) \\
I3a & ---  & --- & 2   & 799 (u)  & (-17.54,-37.16) (-17.54,-18.21) (-17.54,-7.41) \\
    &      &     &     &          & (-5.07,-0.0013)* (-4.49,-0.0015)* (-4.02,-0.0032)* \\
I3b & ---  & --- & --- & 864 (n)  & (-17.54,-37.14) (-17.54,-18.20) (-17.54,-7.40) \\
    &      &     &     &          & (-6.78,-0.0008)* \\
I4a & ---  & --- & 3   & 864 (n)  &  (-26.31,-37.13) (-26.31,-18.60) (-26.31,-7.70) \\
    &      &     &     &          & (-17.16,-0.0029)* (-7.70,-0.0019)* (-5.83,-0.0035)* (-5.39,-0.0019) \\
I4b & ---  & --- & --- & 1301 (n) & (-26.31,-37.13) (-26.31,-18.60) (-26.31,-7.70) \\
    &      &     &     &          & (-11.26,-0.0013)* (-9.17,-0.0012)* (-6.53,-0.0026)* (-6.30,-0.0022)* \\ 
    &      &     &     &          & (-5.83,-0.0024)* (-5.40,-0.0037) \\
I5  & ---  & 0.04& 1   & 799 (u)  & (-9.00,-37.07) (-9.00,-17.60) (-9.00,-6.88) \\
II  & 0.06 & 0.1 & 1   & 799 (u)  & (-5.93,-40.33) (-5.93,-28.16) (-5.93,-23.06) \\
    &      &     &     &          & (-5.93,-17.41) (-5.93,-12.95) (-5.93,-6.91)\\

\hline
\end{tabular}
\label{table1}
\end{table}

\no The grids corresponding to $n_r=864,$ $1175$ and $1301$ are non-uniform. 
They are characterised by a step $\Delta x_1$ between $x=0.1$
and $0.3$, $\Delta x_2$ between $x=0.3$ and $0.8$, and $\Delta x_3$
between $x=0.8$ and $1$. The values of $\Delta x_1,$ $\Delta x_2$ and
$\Delta x_3$ are given in table~\ref{table2} where $\Delta
x_0=0.9/399$.

\begin{table}
\caption{Characteristics of the non uniform grids}
\begin{tabular} {cccccccccccccccc} \hline 

$n_r$ & $\Delta x_1$ & $\Delta x_2$ & $\Delta x_3$ \\ 
\hline

864  & $\Delta x_0/2$ & $\Delta x_0$ & $\Delta x_0/4$ \\
1175 & $\Delta x_0/8$ & $\Delta x_0/2$ & $\Delta x_0/2$ \\
1301 & $\Delta x_0/2$ & $\Delta x_0$ & $\Delta x_0/8$ \\

\hline
\end{tabular}
\label{table2}
\end{table}

\begin{figure}
\centerline{\epsfig{file=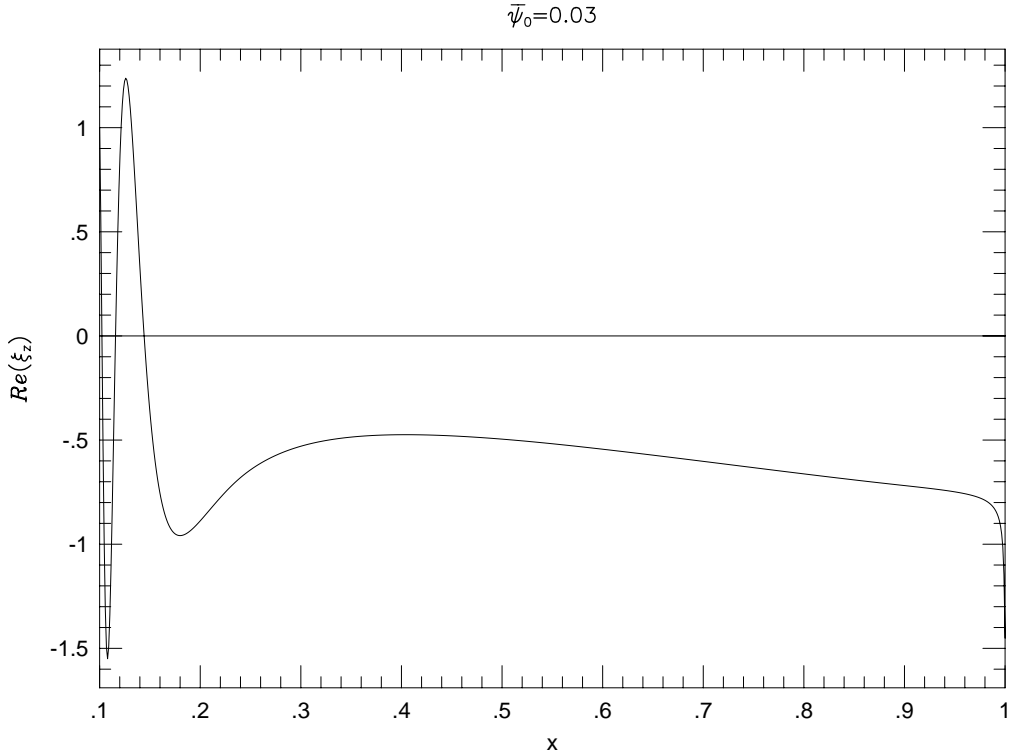,height=120mm,angle=0}}
\caption[]{${\rm Re}(\xi_z)$ for $\bar{\psi}_0=0.03$, $\beta_0=0.1$, $m=1$
and $n_r=799$ (model~I2a). The mode represented is the modified rigid
tilt mode and has $\bar{\sigma}=(-0.025,0)$.}
\label{fig_tilt}
\end{figure}

\no We performed a test for model~I2a by setting $\psi_{\rm {ext}} =0.$ In
this case with no external field, we confirmed that the disc has a
zero frequency rigid tilt mode ($\xi_z \propto r$) which is the mode
having the lowest frequency. When an external field is added, the
equivalent mode is a modified rigid tilt mode, as shown on
Fig.~\ref{fig_tilt}. Since the dipole field is significant only in
the inner parts of the disc, this mode takes on the character of a
rigid tilt mode in the outer parts.

\no For $m=0$ and $m=1,$ all the unstable modes are well resolved. For
higher values of $m,$ in addition to well resolved unstable modes we
get some poorly resolved weakly unstable ones. The number, frequency
and growth rate of these modes depend on the grid resolution. However,
in all cases they have growth rates several orders of magnitude
smaller than those of the well resolved modes. Thus, although their
reality is questionable, they are not important, and from now on we
shall consider only the spectrum consisting of the well resolved
unstable modes.

\no These modes are confined in the inner parts of the disc where
$\bar{\kappa}^2 < 0$ (see Fig.~\ref{figd}). This is in agreement
with the local dispersion relation which predicts that instability
ensues when the condition~(\ref{cond}) is satisfied. The frequency of
these modes is $-m \bar{\Omega}_{\rm c}$, where $\bar{\Omega}_{\rm c}$ is the
angular velocity in the inner parts of the disc, and their growth rate
depends only weakly on $m$. For each $m,$ the most unstable modes have
a growth rate significantly larger than their frequency indicating
dynamical instability. The number of modes in the unstable spectrum
increases with the strength of the magnetic support in the inner parts
of the disc, as does the growth rate of the most unstable mode.

\begin{figure}
\centerline{\epsfig{file=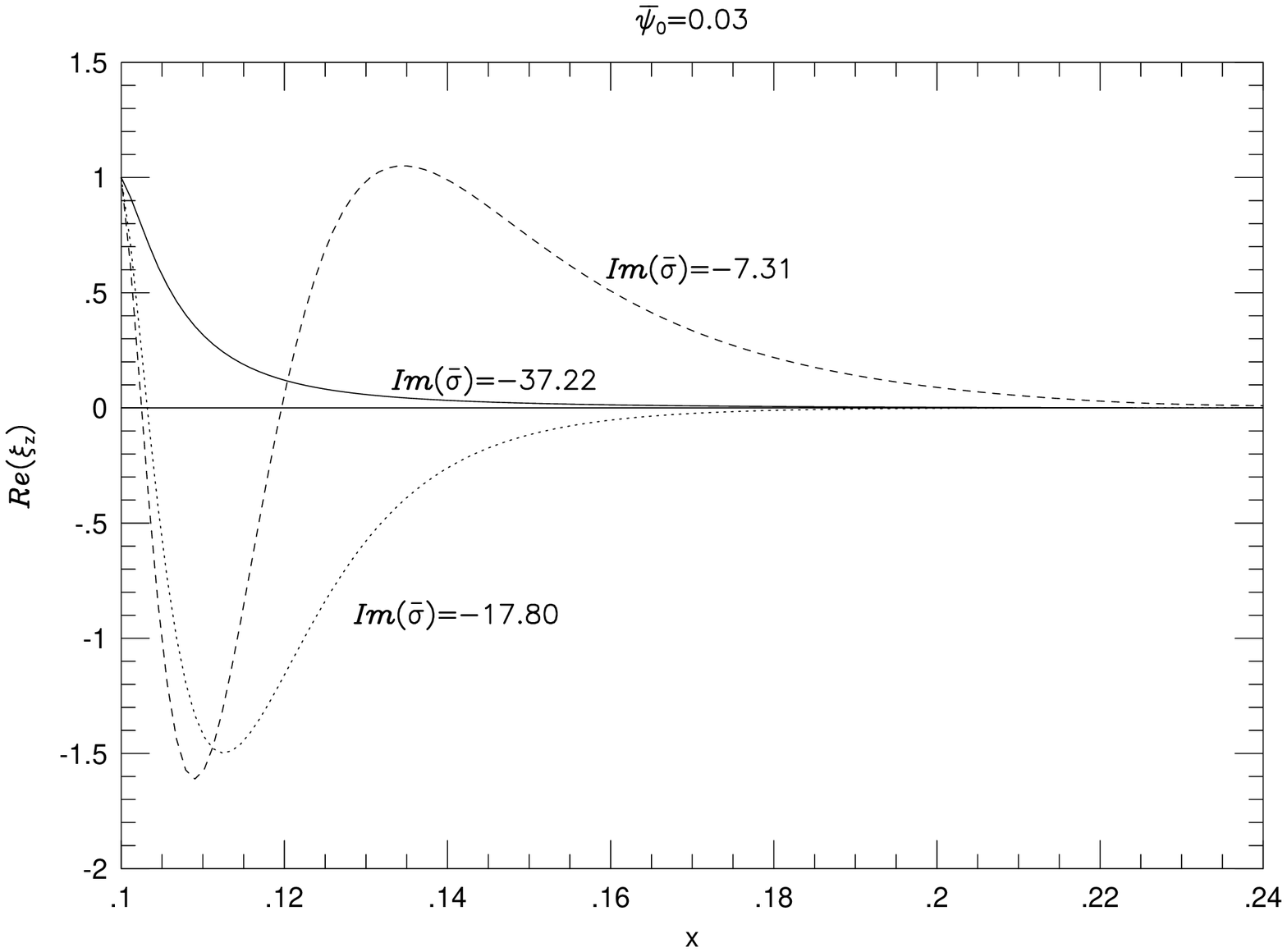,height=120mm,angle=0}}
\caption[]{${\rm Re}(\xi_z)$ for $\bar{\psi}_0=0.03$, $\beta_0=0.1$, $m=1$
and $n_r=799$ (model~I2a). The modes represented have
${\rm Re}(\bar{\sigma})=-8.77$ and ${\rm Im}(\bar{\sigma})=-37.22$ (solid line),
${\rm Im}(\bar{\sigma})=-17.80$ (dotted line) and ${\rm Im}(\bar{\sigma})=-7.31$
(dashed line). Only the inner parts of the disc are represented.}
\label{fig1}
\end{figure}

\no For $\bar{\psi}_0=0.03$ there are 3 unstable modes, with
respectively 0, 1 and 2 nodes in the real part of $\xi_z,$ whereas for
$\bar{\psi}_0=0.06$ there are 6 modes, with the number of nodes
varying between 0 and 5 (the smaller the growth rate, the larger the number
 of nodes). Fig.~\ref{fig1} and Fig.~\ref{fig2} show
the real part of $\xi_z$ in the inner parts of the disc for models~I2a
and~II respectively. In all cases the imaginary part of $\xi_z$ is
very small compared to its real part. We note that the characteristics
of these modes do not depend on the resolution.

\begin{figure}
\centerline{\epsfig{file=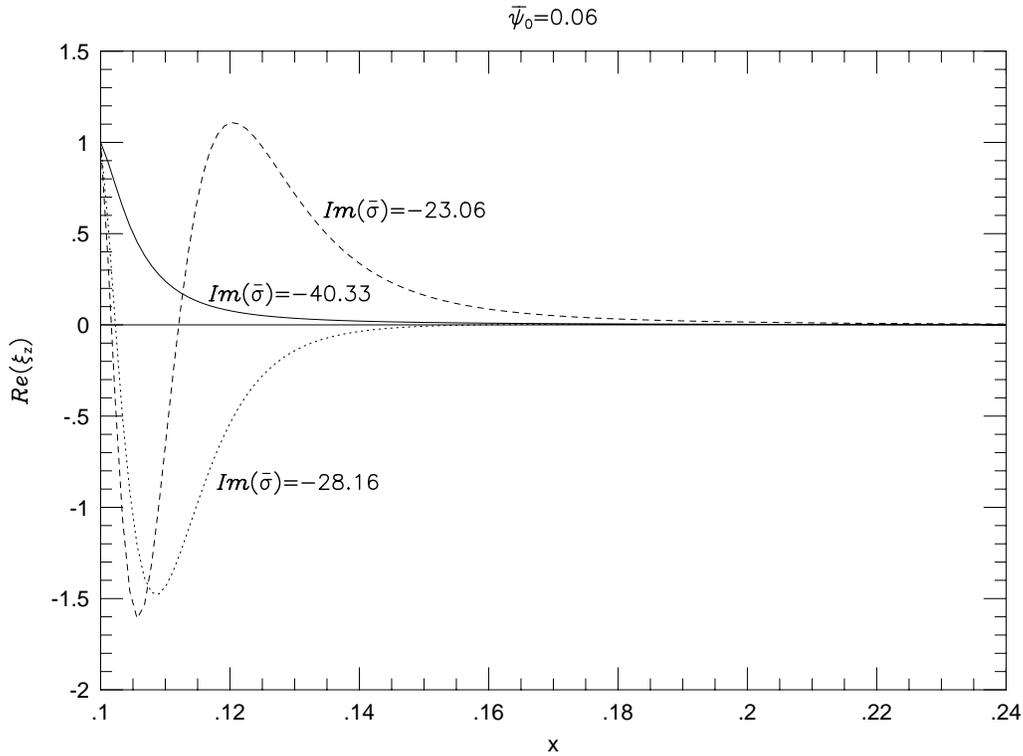,height=120mm,angle=0}}
\caption{${\rm Re}(\xi_z)$ for $\bar{\psi}_0=0.06$, $\beta_0=0.1$,
$m=1$ and $n_r=799$ (model~II). The modes represented have
${\rm Re}(\bar{\sigma})=-5.93$ and ${\rm Im}(\bar{\sigma})=-40.33$ (solid line),
${\rm Im}(\bar{\sigma})=-28.16$ (dotted line) and ${\rm Im}(\bar{\sigma})=-23.06$
(dashed line). Only the inner parts of the disc are represented.}
\label{fig2}
\end{figure}

\begin{figure}
\centerline{\epsfig{file=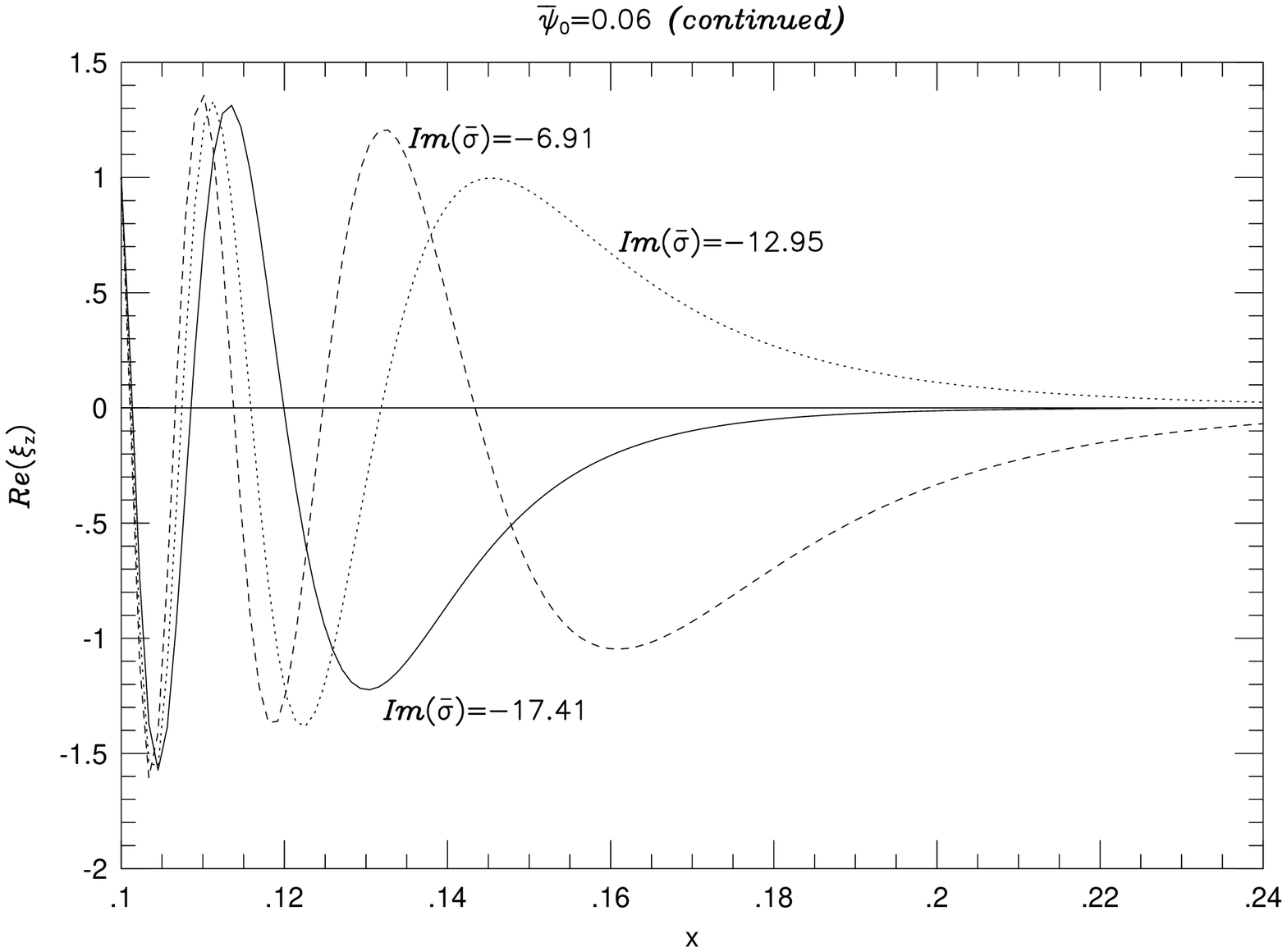,height=120mm,angle=0}}
\contcaption{ The modes represented have
${\rm Re}(\bar{\sigma})=-5.93$ and ${\rm Im}(\bar{\sigma})=-17.41$ (solid line),
${\rm Im}(\bar{\sigma})=-12.95$ (dotted line) and ${\rm Im}(\bar{\sigma})=-6.91$
(dashed line).}
\end{figure}

\no Varying $\beta_0$ hardly changes the characteristics of the unstable
modes. As mentioned above, the inner parts of the disc, where the
modes are confined, are indeed dominated by the dipole field where
conditions are insensitive to $\beta_0.$

\no In the equilibrium models to which the above calculations
correspond, the internal vertical field vanishes at some location in
the outer parts of the disc. To check whether the results depend on
the unlikely presence of this O-point, we have re-run model~I2a and
model~I3 with $n_r=1301$ and the outer boundary at $x=0.7$. The
results are very similar to those obtained above, the minor
differences coming from the fact that the rotation and the density
profiles are a bit different in the two cases. This is consistent with
the fact that the internally confined unstable modes are almost
independent of the structure of the outer parts of the disc.

\section{Discussion}

\label{sec:discussion}

In this paper, we have studied the stability of a magnetized accretion
disc to disturbances perpendicular to its plane (bending modes).
At equilibrium, the disc is permeated by both an internally
produced poloidal magnetic field and an external dipole field. The
former arises from a toroidal current in the disc, the latter is
supposed to originate from a magnetized central star. In the important
inner regions of the disc, the field lines are in a state of
isorotation. We suppose that the uniformly rotating inner disc could
be produced by material diffusing inwards through the action of
interchange instabilities (Spruit~\& Taam~1990) when the mass
injection rate into the disc is high enough to enable that to occur
with ultimate accretion onto and spin up of the central star. We have
neglected a possible toroidal component of the magnetic field which
would be produced through differential rotation if poloidal field
lines connected points with different angular velocity. Since the disc
is supposed to be infinitesimally thin, radial pressure gradients have
been neglected.

\no A local stability analysis leads to the $m$ independent condition
for instability $\kappa^2_{\rm m} < 0$ (see~(\ref{cond})), where
$\kappa_{\rm m}$
is the modified epicyclic frequency defined by the
relation~(\ref{kappa}). In the inner regions of the disc, where the
dipole field predominates over the internal one, this condition is
satisfied if the magnetic field provides enough support against
gravity (see~(\ref{cond1})). We note that even though uniform rotation
acts in favour of the instability, bending modes in an entirely
differentially rotating disc may also be unstable. An external dipole
is not absolutely necessary for instability to occur.

\no For axisymmetric modes, the existence of a variational principle
leads to the rigorous global criterion~(\ref{var}) which is equivalent
to~(\ref{cond}) when local displacements are considered. Even though
the razor--thin disc approximation has been used so far, the
instabilities are not an {\em ab initio} artefact of this assumption. The
criterion~(\ref{var}) is the thin disc limit of the general
variational principle of Papaloizou~\& Szuszkiewicz~(1992) for
axisymmetric modes.

\no We have solved numerically the normal mode equation~(\ref{mod})
for specific disc equilibrium models. For low values of $m$ we get a
spectrum of well resolved dynamically unstable modes, which are
confined in the inner parts of the disc where $\kappa^2_{\rm m} < 0$. For a
given $m,$ the number $N$ of these modes increases with the magnetic
support. For these the tendency is that the number of nodes in the
real part of the vertical displacement increases from 0 to $N-1$. This
is similar to what happens in a non-magnetized, self-gravitating,
uniformly rotating disc (we have already mentioned in
section~\ref{sec:dispersion} the similarity between the two
situations). In that case, Hunter~\& Toomre~(1969) have indeed shown
that the normal modes of free oscillation are polynomials with
increasing numbers of nodes. However, in the self-gravitating disc,
the bending modes are always stable.

\no We have found numerically that the pattern speed associated with
the unstable modes is $\sim \Omega_{\rm c}$, with $\Omega_{\rm c}$ being the
constant angular velocity in the inner parts of the disc.

\no We have argued that the instability of axisymmetric modes can be
thought of as an unstable interaction between the disc current and the
central dipole. Instability arises from the fact that the energy of
the dipole in the magnetic field produced by the disc currents
decreases due to the perturbation.

\no The determination of the outcome of these instabilities awaits a 
non-linear analysis. However, we comment that if such instabilities
occur, a configuration where the inner regions of the disc are
displaced from the equatorial plane of the central star may be
possible. Then the accretion of the disc material along the dipole
field lines will be facilitated and may occur preferentially onto one
stellar hemisphere depending on the mixture of normal modes
present. Non-axisymmetric instabilities may result in a quasi periodic
light variation even when the disc angular momentum vector at large
distances and the central dipole axis are aligned. The asymmetric
magnetospheric accretion would result in an observational signature of
the star-disc configuration. The periodic light variation of Classical
T~Tauri stars is mostly interpreted as rotational modulation of the
stellar flux by hot spots due to magnetospheric accretion (see, for
example, Bouvier et al.~1995). If accretion occurs preferentially to
one hemisphere, then depending on the orientation of the observer the
periodic light variation will be observed.
The non-axisymmetric bending of the disc plane would
lead to hot spots being created on the stellar surface and to the
rotational modulation of the stellar output, even when the axis of the
stellar dipole is aligned with the axis of rotation of the accretion disc.

\section*{Acknowledgments}
This work was supported by PPARC grant GR/H/09454 and the EU grant
ERB-CHRX-CT93-0329. C.T. acknowledges support by the Center for Star
Formation Studies at NASA--Ames Research Center and the University of
California at Berkeley and Santa-Cruz. V.A. acknowledges support by the
State Scholarships Foundation (IKY) of the Republic of Greece through
a postgraduate studentship.


\begin{thebibliography}{}

\bibitem[1996]{Agapitou}
Agapitou V., Papaloizou J.C.B., 1996, Astro. Lett. \& Comm., 34, 363

\bibitem[1969]{Anzer}
Anzer U., 1969, Solar Physics, 8, 37

\bibitem[1996]{Armitage}
Armitage P.J., Clarke C.J., 1996, MNRAS, 280, 458

\bibitem[1994]{Bouvier94}
Bouvier J., 1994, in Caillault J.-P., ed.,Eighth Cambridge Workshop on
Cool Stars, Stellar Systems and the Sun, ASP Conference Series, vol. 64

\bibitem[1995]{Bouvier95}
Bouvier J., Covino E., Kovo O., Martin E.L., Matthews J.M., Terranegra
L., Beck S.C., 1995, A\&A, 299, 89

\bibitem[1989]{brand}
Brandenburg A., Tuominen I., Moss D., 1989, Geophys. Astrophys. Fluid
Dyn., 49, 129

\bibitem[1990]{Cabrit}
Cabrit S., Edwards S., Strom S.E., Strom K.M., 1990, ApJ, 354, 687

\bibitem[1992]{Calvet}
Calvet N., Hartmann L., 1992, ApJ, 386, 239

\bibitem[1990]{Camenzind}
Camenzind M., 1990, Rev. Mod. Astron., 3, 234

\bibitem[1993]{Cameron}
Cameron A.C., Campbell C.G., 1993, MNRAS, 274, 309

\bibitem[1987]{Campbell}
Campbell C.G., 1987, MNRAS, 229, 405

\bibitem[1994]{Edwardsa}
Edwards S., Hartigan P., Ghandour L., Androulis C., 1994,
AJ, 108, 1056

\bibitem[1993]{Edwardsab}
Edwards S., Strom S.E., Hartigan P., Strom K.M., Hillenbrand L.A., 1993,
AJ, 106, 372

\bibitem[1995]{Ghosh1}
Ghosh P., 1995, MNRAS, 272, 763

\bibitem[1977]{Ghosh2}
Ghosh P., Lamb F.K., 1978, ApJ, 223, L83

\bibitem[1979]{Ghosh3}
Ghosh P., Lamb F.K., 1979, ApJ, 234, 296 

\bibitem[1994]{Hartmann}
Hartmann L., Hewett R., Calvet N., 1994, ApJ, 426, 669

\bibitem[1969]{Hunter}
Hunter C., Toomre A., 1969, ApJ, 155, 747

\bibitem[1975]{Jackson}
Jackson J.D., 1975, Classical Electrodynamics. Wiley, New York

\bibitem[1996]{Kenyon}
Kenyon S.J., Yi I., Hartmann L., 1996, ApJ, 462,439

\bibitem[1989]{Ko89}
K\"onigl A., 1989, ApJ, 342, 208

\bibitem[1991]{Ko91}
K\"onigl A., 1991, ApJ, 370, L39

\bibitem[1993]{Ko93}
K\"onigl A., 1993, in Levy E.H., Lunine J.I., eds., Protostars and
Planets III. Univ. Arizona Press, Tuscon, p. 641

\bibitem[1995]{Lehmann}
Lehmann T., Reipurth B., Brandner W., 1995, A\&A, 300, L9

\bibitem[1996]{Lepeltier}
Lepeltier T., Aly J.J., 1996, A\&A, 306, 645 

\bibitem[1994]{Lubow}
Lubow S.H., Papaloizou J.C.B., Pringle J.E., 1994, MNRAS, 267, 235 

\bibitem[1993]{Montmerle1}
Montmerle T., Andr'e P., Casanova S., Feigelson E.D., 1993, in
Lynden-Bell D., ed., NATO Advanced Worksop on Cosmical
Magnetism. Kluwer, Dordrecht

\bibitem[1994]{Montmerle2}
Montmerle T., Feigelson E.D., Bouvier J., Andr'e P., 1994, in Levy
E.H., Lunine J.I., eds., Protostars and Planets III. Univ. Arizona
Press, Tuscon, p. 689

\bibitem[1997]{Ogilvie}
Ogilvie G., 1997, MNRAS, 288, 63

\bibitem[1996]{Paatz}
Paatz G., Camenzind M., 1996, A\&A, 308, 77

\bibitem[1991]{Paczynski}
Paczynski B., 1991, ApJ, 370, 597

\bibitem[1995]{PapaloizouL}
Papaloizou J.C.B., Lin D.N.C., 1995, ARA\&A, 33, 505 

\bibitem[1992]{Papaloizou}
Papaloizou J.C.B., Szuszkiewicz E., 1992, Geophys. Astrophys. Fluid
Dyn., 66, 223 

\bibitem[1995]{PapaloizouT}
Papaloizou J.C.B., Terquem C., 1995, MNRAS, 227,553

\bibitem[1991]{Popham}
Popham R., Narayan R., 1991, ApJ, 370, 604

\bibitem[1986]{Press}
Press W.H., Flannery B.P., Teukolsky S.A., Vetterling W.T., 1986, 
Numerical Recipes: The Art of Scientific Computing, Cambridge 
University Press 

\bibitem[1996]{Reyez-Ruiz}
Reyez-Ruiz M., Stepinski T.F., 1996, ApJ, 459, 653 

\bibitem[1990]{Shu} 
Shu F.H., 1984, in Greenberg R., Brahic A., eds., Planetary Rings.
Univ. Arizona Press, Tuscon, p. 513 

\bibitem[1990]{Spruit}
Spruit H.C., Taam R.E., 1990, A\&A, 229, 475 

\bibitem[1993]{Spruit3}
Spruit H.C., Taam R.E., 1993, ApJ, 402, 593 

\bibitem[1993]{Spruit4}
Spruit H.C., Stehle R., Papaloizou J.C.B., 1995, MNRAS, 275, 1223 

\bibitem[1990]{Tagger}
Tagger M., Henriksen R.N., Sygnet J.F., Pellat R., 1990, ApJ, 353, 654

\bibitem[1987]{Tayler}
Tayler R. J., 1987, MNRAS, 227, 553
\bibitem[1987]{Wu}
Wu F., 1987, ApJ, 320, 418 

\bibitem[1994]{Yi}
Yi I., 1994, ApJ, 428, 760

\label{lastpage}
\end{thebibliography}
\end{document}